\documentclass[preprint,aps,prd]{revtex4}
\usepackage{epsfig}
\usepackage{graphicx}
\usepackage{dcolumn}
\usepackage{bm}

\date{}
\begin{document}


\newcommand{\ds}{\displaystyle}
\newcommand{\mc}{\multicolumn}
\newcommand{\bce}{\begin{center}}
\newcommand{\ece}{\end{center}}
\newcommand{\beq}{\begin{equation}}
\newcommand{\eeq}{\end{equation}}
\newcommand{\bea}{\begin{eqnarray}}

\newcommand{\eea}{\end{eqnarray}}
\newcommand{\cont}{\nonumber\eea\bea}
\newcommand{\cl}[1]{\begin{center} {#1} \end{center}}
\newcommand{\ba}{\begin{array}}
\newcommand{\ea}{\end{array}}

\newcommand{\ab}{{\alpha\beta}}
\newcommand{\cd}{{\gamma\delta}}
\newcommand{\dc}{{\delta\gamma}}
\newcommand{\ac}{{\alpha\gamma}}
\newcommand{\bd}{{\beta\delta}}
\newcommand{\abc}{{\alpha\beta\gamma}}
\newcommand{\eps}{{\epsilon}}
\newcommand{\lam}{{\lambda}}
\newcommand{\mn}{{\mu\nu}}
\newcommand{\mpnp}{{\mu'\nu'}}
\newcommand{\Amuu}{{A_{\mu}}}
\newcommand{\Amuo}{{A^{\mu}}}
\newcommand{\Vmuu}{{V_{\mu}}}
\newcommand{\Vmuo}{{V^{\mu}}}
\newcommand{\Anuu}{{A_{\nu}}}
\newcommand{\Anuo}{{A^{\nu}}}
\newcommand{\Vnuu}{{V_{\nu}}}
\newcommand{\Vnuo}{{V^{\nu}}}
\newcommand{\Fmnu}{{F_{\mu\nu}}}
\newcommand{\Fmno}{{F^{\mu\nu}}}

\newcommand{\abcd}{{\alpha\beta\gamma\delta}}


\newcommand{\bsigma}{\mbox{\boldmath $\sigma$}}
\newcommand{\btau}{\mbox{\boldmath $\tau$}}
\newcommand{\brho}{\mbox{\boldmath $\rho$}}
\newcommand{\bpipi}{\mbox{\boldmath $\pi\pi$}}
\newcommand{\bss}{\bsigma\!\cdot\!\bsigma}
\newcommand{\btt}{\btau\!\cdot\!\btau}
\newcommand{\bnabla}{\mbox{\boldmath $\nabla$}}
\newcommand{\bphi}{\mbox{\boldmath $\tau$}}
\newcommand{\bvarphi}{\mbox{\boldmath $\rho$}}
\newcommand{\bDelta}{\mbox{\boldmath $\Delta$}}
\newcommand{\bpsi}{\mbox{\boldmath $\psi$}}
\newcommand{\bPsi}{\mbox{\boldmath $\Psi$}}
\newcommand{\bPhi}{\mbox{\boldmath $\Phi$}}
\newcommand{\bnab}{\mbox{\boldmath $\nabla$}}
\newcommand{\bpi}{\mbox{\boldmath $\pi$}}
\newcommand{\btheta}{\mbox{\boldmath $\theta$}}
\newcommand{\bkappa}{\mbox{\boldmath $\kappa$}}

\newcommand{\bA}{{\bf A}}
\newcommand{\bB}{\mbox{\boldmath $B$}}
\newcommand{\bp}{\mbox{\boldmath $p$}}
\newcommand{\bk}{\mbox{\boldmath $k$}}
\newcommand{\bq}{\mbox{\boldmath $q$}}
\newcommand{\bfe}{{\bf e}}
\newcommand{\bb}{\mbox{\boldmath $b$}}
\newcommand{\br}{\mbox{\boldmath $r$}}
\newcommand{\bR}{\mbox{\boldmath $R$}}

\newcommand{\fph}{${\cal F}$}
\newcommand{\aph}{${\cal A}$}
\newcommand{\dph}{${\cal D}$}
\newcommand{\fpi}{f_\pi}
\newcommand{\mpi}{m_\pi}
\newcommand{\Tr}{{\mbox{\rm Tr}}}
\def\Qb{\overline{Q}}
\newcommand{\delu}{\partial_{\mu}}
\newcommand{\delo}{\partial^{\mu}}
%
%
\newcommand{\up}{\!\uparrow}
\newcommand{\upup}{\uparrow\uparrow}
\newcommand{\updo}{\uparrow\downarrow}
\newcommand{\uu}{$\uparrow\uparrow$}
\newcommand{\ud}{$\uparrow\downarrow$}
\newcommand{\auu}{$a^{\uparrow\uparrow}$}
\newcommand{\aud}{$a^{\uparrow\downarrow}$}
\newcommand{\pu}{p\!\uparrow}

\newcommand{\qp}{quasiparticle}
\newcommand{\sa}{scattering amplitude}
\newcommand{\ph}{particle-hole}
\newcommand{\qcd}{{\it QCD}}
\newcommand{\integ}{\int\!d}
\newcommand{\ie}{{\sl i.e.~}}
\newcommand{\etal}{{\sl et al.~}}
\newcommand{\etc}{{\sl etc.~}}
\newcommand{\rhs}{{\sl rhs~}}
\newcommand{\lhs}{{\sl lhs~}}
\newcommand{\eg}{{\sl e.g.~}}
\newcommand{\ef}{\epsilon_F}
\newcommand{\sigt}{d^2\sigma/d\Omega dE}
\newcommand{\sige}{{d^2\sigma\over d\Omega dE}}
\newcommand{\rpaeq}{\beq
\left ( \begin{array}{cc}
A&B\\
-B^*&-A^*\end{array}\right )
\left ( \begin{array}{c}
X^{(\kappa})\\Y^{(\kappa)}\end{array}\right )=E_\kappa
\left ( \begin{array}{c}
X^{(\kappa})\\Y^{(\kappa)}\end{array}\right )
\eeq}
\newcommand{\ket}[1]{| {#1} \rangle}
\newcommand{\bra}[1]{\langle {#1} |}
\newcommand{\ave}[1]{\langle {#1} \rangle}
\newcommand{\half}{{1\over 2}}

\newcommand{\singlespace}{
    \renewcommand{\baselinestretch}{1}\large\normalsize}
\newcommand{\doublespace}{
    \renewcommand{\baselinestretch}{1.6}\large\normalsize}
\newcommand{\bftau}{\mbox{\boldmath $\tau$}}
\newcommand{\bfalpha}{\mbox{\boldmath $\alpha$}}
\newcommand{\bfgamma}{\mbox{\boldmath $\gamma$}}
\newcommand{\bfxi}{\mbox{\boldmath $\xi$}}
\newcommand{\bfbeta}{\mbox{\boldmath $\beta$}}
\newcommand{\bfeta}{\mbox{\boldmath $\eta$}}
\newcommand{\bfpi}{\mbox{\boldmath $\pi$}}
\newcommand{\bfphi}{\mbox{\boldmath $\phi$}}
\newcommand{\bfR}{\mbox{\boldmath ${\cal R}$}}
\newcommand{\bfL}{\mbox{\boldmath ${\cal L}$}}
\newcommand{\bfM}{\mbox{\boldmath ${\cal M}$}}
\def\dblint{\mathop{\rlap{\hbox{$\displaystyle\!\int\!\!\!\!\!\int$}}
    \hbox{$\bigcirc$}}}
\def\ut#1{$\underline{\smash{\vphantom{y}\hbox{#1}}}$}

\def\UNITY{{\bf 1\! |}}
\def\Pom{{\bf I\!P}}
\def\lsim{\mathrel{\rlap{\lower4pt\hbox{\hskip1pt$\sim$}}
    \raise1pt\hbox{$<$}}}         
\def\gsim{\mathrel{\rlap{\lower4pt\hbox{\hskip1pt$\sim$}}
    \raise1pt\hbox{$>$}}}         
\def\beq{\begin{equation}}
\def\eeq{\end{equation}}
\def\bea{\begin{eqnarray}}
\def\eea{\end{eqnarray}}

\doublespace
{\large ~~~~~~~~~~~~~~~~~~~~~~~~~~~~~~~~~~~~~~~~~~~~~~~~~~~~~~~~~~FZJ-IKP-TH-2004-21}\\
        
\begin{center}

{\Large\bf Breaking of $k_{\perp}$-factorization for Single Jet
Production off Nuclei}\\ \vspace{1cm}
 { \bf N.N. Nikolaev$^{a,b)}$,
W. Sch\"afer$^{a)}$\medskip\\  }

\vspace{1.0cm} {\sl $^{a}$IKP(Theorie), Forschungszentrum
J{\"u}lich, D-52425 J{\"u}lich, Germany
\medskip\\
$^{b}$L.D. Landau Institute for Theoretical Physics, Moscow
117940, Russia}
\medskip\\
E-mail: N.Nikolaev$@$fz-juelich.de; Wo.Schaefer$@$fz-juelich.de\vspace{1cm} \\
{\bf Abstract\\    }
\end{center}

{\small The linear $k_{\perp}$-factorization is part and parcel of
the perturbative QCD description of high energy hard processes off free nucleons.
 In the case of heavy nuclear targets
the very concept of nuclear
parton density becomes ill-defined as exemplified by the recent
derivation \cite{Nonlinear} of nonlinear nuclear
$k_{\perp}$-factorization for forward dijet
production in deep inelastic scattering 
off nuclei. Here we report a derivation of the
related breaking of  $k_{\perp}$-factorization for single-jet
processes. We present a general formalism and apply it to several
cases of practical interest: open charm and quark \& gluon jet
production in
the central to beam fragmentation region of $\gamma^*p,\gamma^*A, pp$
and $pA$ collisions. We show how the pattern of $k_{\perp}$-factorization
breaking and the nature and number of exchanged nuclear pomerons 
do change within the phase space of produced quark and gluon jets.
As an application of the
nonlinear  $k_{\perp}$-factorization we discuss the Cronin effect.
Our results are also applicable to the $p_{\perp}$-dependence of the
Landau-Pomeranchuk-Migdal effect for, and nuclear quenching of,
jets produced in the proton hemisphere
of $pA$ collisions.}

\begin{center}
\end{center}

\pagebreak


\section{Introduction}
\label{Sec:1}

The familiar perturbative QCD (pQCD) factorization theorems tell 
that the hard scattering
observables are linear functionals (convolutions) of the appropriate
parton densities
in the projectile and target \cite{Textbook} and are based on an implicit
assumption that in the low-parton density limit the unitarity constraint
can be ignored. They play, though, a fundamental role in interaction with
strongly absorbing heavy nuclei which puts into question the very concept
of a nuclear parton density. Indeed, a consistent analysis of forward hard
dijet production in deep inelastic scattering (DIS)
off nuclei revealed a striking breaking of 
$k_{\perp}$-factorization \cite{Nonlinear}: Namely, starting with the
nuclear-shadowed nuclear DIS cross section, one can define the collective
nuclear unintegrated gluon density such that the familiar linear
$k_{\perp}$-factorization (see e.g. the recent reviews \cite{Andersson:2002cf})
would hold for the forward single-quark spectrum
but fail for the quark-antiquark two-jet spectrum which turns out to be
a highly nonlinear functional of the collective nuclear gluon density. This
finding casts a shadow on numerous extensions to nuclear collisions of the
parton model formalism developed for nucleon-nucleon collisions and
calls upon for revisiting other hard processes. The ultimate goal is 
the consistent description of nuclear effects from DIS to 
ultrarelativistic hadron-nucleus to (the initial stage of) 
nucleus-nucleus collisions, for instance at RHIC and LHC.

The purpose of this communication is a demonstration that the pattern of
$k_{\perp}$-factorization breaking discovered in \cite{Nonlinear} holds
also for a broad class of single-jet spectra. Here we recall that nuclear
shadowing in DIS sets in at the Bjorken variable
\beq
x\lsim x_A={1\over R_A m_N}\,,
\label{eq:1.1}
\eeq
when the coherency over the thickness of the nucleus holds for the
$q\bar{q}$ Fock states of the virtual photon (\cite{NZfusion,NZ91},
for the color dipole phenomenology of the
experimental data on nuclear shadowing see \cite{BaroneShadowing}). Here $R_A$ is
the radius of the target nucleus of mass number $A$ and $m_N$ is the proton mass. The analysis
\cite{Nonlinear} focused on the excitation of quark and antiquark
jets, $\gamma^* \to q\bar{q}$, in the photon fragmentation region
at $x\lsim x_A$. The subject of this communication is excitation of
open charm and jets,
\beq
g^* \to Q\bar{Q},\quad g^* \to g g,\quad q^* \to q g,
\label{eq:1.2}
\eeq
in  hadron--nucleon/nucleus scattering and/or 
DIS off nucleons and nuclei at $x \ll x_A$. At RHIC that means 
the large (pseudo)rapidity region of proton-proton and
proton-nucleus collisions. 
Our starting point is the
color dipole multiple scattering theory
formulation of the open charm production in $pA$ collisions
\cite{NPZcharm} which we extend from the total to differential
cross section and to arbitrary color state of the projectile parton.
We treat the jet production at the  pQCD parton level, our approach
to the jet spectra is in many respects
similar to Zakharov's lightcone description of the QCD 
Landau-Pomeranchuk-Migdal (LPM) effect for 
the limit of
thin targets \cite{Slava,SlavaReview}.
The results obtained below
for the open charm and quark and gluon jets 
make manifest a change from linear to nonlinear $k_{\perp}$-factorization
as the target changes from a free nucleon to heavy nucleus. 

In the production of dijets in DIS \cite{Nonlinear} and $\pi A$ collisions
\cite{PionDijet} the fragmenting particle was a color-singlet one. Here
we consider the $gN,gA,qN,qA$ collisions in which the beam partons
are colored. 
In general the scattering of colored particles is plagued
by infrared problems, but the 
excitation cross sections can be calculated as combinations of
infrared-safe total cross sections for interaction with the target of certain
color singlet multiparton states \cite{Nonlinear,NPZcharm,Slava}. 
We discuss to some detail the so-called Cronin effect \cite{Cronin} - the 
nuclear antishadowing at moderate to large transverse momentum $\bp$ of jets. 
Decomposing in the spirit of the Leading Log$\bp^2$ (LL$\bp^2$) 
approximation the single-jet cross section into the so-called
direct and resolved interactions of the incident parton we demonstrate
that the Cronin effect is a salient feature of the resolved interactions
and  identify its origin to be the antishadowing nuclear higher-twist
component of the collective gluon density of a nucleus derived in 
\cite{NSSdijet,NSSdijetJETPLett}. 
Regarding the potential applications to $pA$ collisions at RHIC
we note that the $k_{\perp}$-factorization phenomenology of DIS 
\cite{INDiffGlue} suggests fairly moderate $Q_A \sim 1$ GeV at $x\sim 10^{-2}$
for the heaviest nuclei \cite{Nonlinear}, but we find that interesting
nuclear effects extend to the pQCD domain of 
jet transverse momenta way beyond $Q_A$.
Qualitatively similar
results would be expected in a regime of strong absorptive corrections 
also for hadronic targets other than nuclei, the nucleus 
represents a theoretical laboratory, in which saturation effects are endowed
with a large parameter, the nuclear opacity, which allows for their
systematic account.

One more point needs an emphasis. The partonic cross
sections evaluated within the often used collinear approximation are
subject to large smearing corrections for the so-called intrinsic 
transverse momentum $\langle p_T\rangle$ of
partons, see \cite{PappIntrinsicKperp}. In contrast to that we treat
the pQCD partonic subprocesses with full allowance for the
transverse momenta of all interacting partons and no extraneous
$\langle p_T\rangle$-smearing is needed. 
The 
discussion of nonperturbative effects such as 
nuclear modifications of the hadronization of jets (\cite{Cassing} 
and references therein)
or of the recombination of partons into high-$p_{\perp}$ hadrons
(\cite{HwaRecombination} and references therein)
goes beyond the scope of the present study. 

All our results are for the distribution of jets in both the transverse 
and longitudinal momentum, the latter is parameterized in terms 
of the fraction, $z$, of the longitudinal momentum of the incident
parton $a$ carried by the observed parton $b$. As such, 
they solve the LPM problem for finite transverse momentum
$ \bp$, while the early works on the LPM effect focused on the
$\bp$-integrated longitudinal momentum spectrum. The 
$\bp$-dependent LPM effect is important for a quantitative
pQCD treatment of the nuclear quenching of forward high-$\bp$
hadrons observed experimentally by the BRAHMS collaboration \cite{BRAHMS}.
The numerical results for the LPM effect for forward jets will be 
reported elsewhere,
here we focus on the derivation and implications of the 
nonlinear $k_{\perp}$-factorization
for the transverse momentum distributions.

The presentation of the main material is organized as follows. Our
starting point is the color-dipole $S$-matrix representation for the 
two-body spectrum from the fragmentation $a\to bc$. We demonstrate
how the single-parton cross spectrum can be calculated in terms of
the $S$-matrices for special 2- and 3-body color singlet parton states
(some technical details are given in Appendix A). The color-dipole
master formulas for the single-parton spectra in interactions with free-nucleon
and heavy-nucleus targets are reported in section \ref{Sec:3}. They form
the basis of the $k_{\perp}$-factorization representation for the
spectra of different partons and in section \ref{Sec:4} we start with the
linear, and in section \ref{Sec:5} with nonlinear, $k_{\perp}$-factorization
for open charm production off free nucleons and nuclei, respectively.
The subject of section \ref{Sec:6} is the Cronin effect in open charm 
production off nuclei. We demonstrate that in the general case
the nuclear spectrum is a quadratic functional
of the collective nuclear gluon density and illustrate this
property in terms of the Kancheli-Mueller diagrams for
inclusive cross sections \cite{KancheliMueller}. 
The Cronin effect is shown to be a generic
feature of the so-called resolved gluon interactions and its
origin is attributed to the 
nuclear antishadowing property of the collective 
nuclear gluon density derived in \cite{NSSdijet}. 
We show that despite the relatively small
saturation scale $Q_A$ the Cronin effect extends to
perturbatively large transverse momenta. 
In section \ref{Sec:7} we apply our master formula to the transverse 
momentum spectrum of gluons from the fragmentation $g^* \to gg$. 
In this case the nuclear cross section is a cubic functional of the
collective nuclear gluon density. 
The subject of section \ref{Sec:8} is the production of quark and gluon
jets in the fragmentation $q^*\to qg$. 
The relevance of our results to the $\bp$-dependent
LPM effect is pointed out in sections \ref{Sec:7} and \ref{Sec:8}. 
The main results are summarized
and discussed in section \ref{Discussion}. Our principal conclusion is that
the $k_{\perp}$-factorization breaking is a universal feature 
of the pQCD description of hard processes in a nuclear environment.
We comment on earlier works on single quark and gluon jet production 
\cite{KovchegovMueller,KopSchafTar,KopNemSchafTar,KharLevMcL,BaierKovWied,
JalilianRaju,Kharzeev:2003wz,Albacete,Accardi,
GelisRaju,Kharzeev:2003sk,Tuchin:2004rb,
Blaizot:2004wu,Iancu:2004bx,Kharzeev:2004yx},
some of which invoke the Color Glass Condensate
approach (\cite{CGC} and references therein).
A convenient technique of derivation
of the $n$-parton interaction S-matrices is presented in Appendix A.
The rules for the derivation of the wavefunctions of two-parton
states in terms of the familiar splitting functions are given in
Appendix B.


\section{Inclusive production as excitation of Fock states of the beam
parton}
\label{Sec:2}

Previously, the breaking of linear $k_{\perp}$-factorization
had been demonstrated for the excitation of dijets, 
$\gamma^* \to q\bar{q}$, in DIS at $x \lsim x_A$ \cite{Nonlinear}. 
Specifically, the two-parton transverse momentum spectrum
was shown to be a highly nonlinear functional of the collective nuclear
gluon density. On the other hand, upon integration over the transverse
momenta of the antiquark jet, the single-quark spectrum was found to
fulfill the linear $k_{\perp}$-factorization. In this paper we extend the
analysis of \cite{Nonlinear} to the case of colored beam partons, 
and confirm the breaking of $k_{\perp}$-factorization as a generic feature 
of the pQCD description of hard processes in a nuclear environment.

To the lowest order in pQCD the underlying subprocess for the mid-
to large-rapidity
open charm
production in proton-proton collisions is the fusion of gluons from
the beam and target, $gg \to c\bar{c}$; for the gluon jet production one
would consider a collision of a beam quark with a gluon from the
target, $qg \to qg$; at higher energies the relevant subprocess for
gluon production will be $gg \to gg$. All the above processes are of the
general form $ag \to bc$ and, from the laboratory frame standpoint, can
be viewed as an excitation of the perturbative $|bc\rangle$ Fock state
of the physical projectile $|a\rangle$ by one-gluon exchange with the
target nucleon. In the case of a nuclear target one has to deal with
multiple gluon exchanges which are enhanced by a large thickness of the
target nucleus. A general treatment of multiple gluon exchanges in
nuclear targets has been developed in \cite{NPZcharm,Nonlinear}, its
extension to the $a\to bc$ transitions can be described as follows:

\begin{figure}[!t]
\begin{center}
\includegraphics[width = 7.0cm, height= 16.0cm,angle=270]{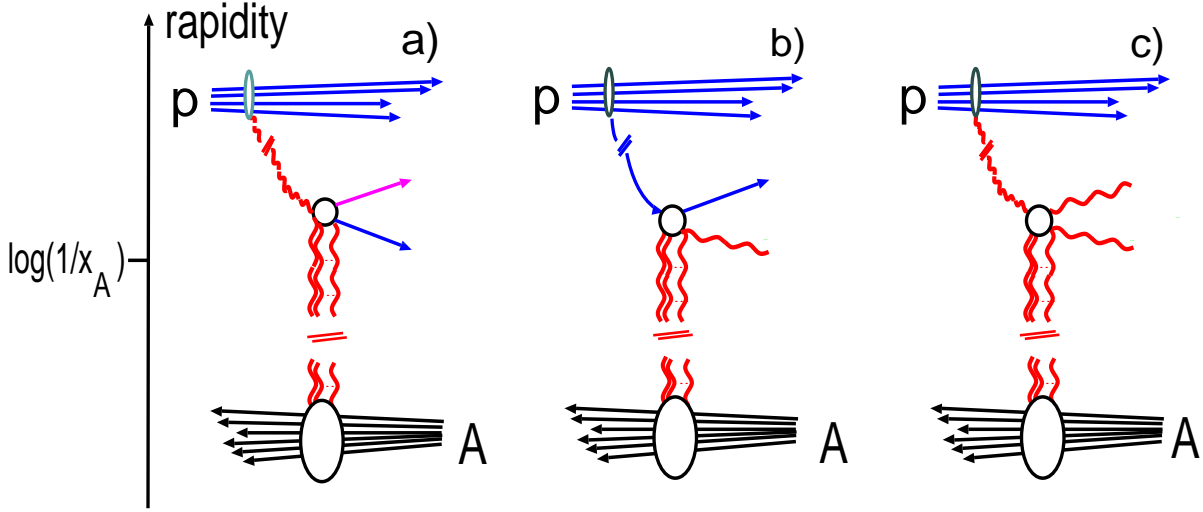}
\caption{
(Color online)
The rapidity structure of the excitation $a \to bc$:
(a) excitation of mid- to large-rapidity open charm $g\to Q\bar{Q}$,
(b) radiation of gluons by quarks $q\to qg$, (c) radiation of gluons
by gluons $g\to gg$.}
\label{fig:SingleJet-Rapidity}
\end{center}
\end{figure}
Partons with energy $E_a$ and transverse momentum
$\bp_{a}$ such that 
\beq
{2m_N E_a  \over \bp_a^2} \gsim {1\over x_A}
\label{eq:2.0}
\eeq 
propagate along straight-line, fixed-impact-parameter, trajectories and 
interact coherently with the nucleus, which is behind the
powerful color dipole formalism \cite{NZ91,NZ92,NZ94}. 
The target frame rapidity structure of the considered $a\to bc$ 
excitation 
is shown in fig.~1. The beam parton has a rapidity $\eta_a >\eta_A=\log{1/x_A}$, 
the final state partons too
have rapidities  $\eta_{b,c} > \eta_A$. In this paper we focus
on the lowest order excitation processes $a\to bc$ without 
production of more secondary partons in the rapidity span
 between $\eta_b$ and $\eta_c$.

To the lowest
order in the perturbative transition $a\to bc$ the Fock state expansion for
the physical state $|a\rangle_{phys}$ reads
\beq
 \ket{a}_{phys} = \ket{a}_0 + \Psi(z_b,\br) \ket{bc}_0
\label{eq:2.1}
\eeq
where $\Psi(z_b,\br)$ is the probability amplitude to find the $bc$ system
with the separation $\br$ in the two-dimensional impact parameter space,
the subscript $"0"$ refers to bare partons. 
The perturbative
coupling of the $a\to bc$ transition is reabsorbed into the lightcone
wave function $\Psi(z_b,\br)$, 
and we also omitted a wave function renormalization factor,
which is of no relevance for the inelastic excitation 
to the perturbative order discussed here. 

\begin{figure}[!t]
\begin{center}
\includegraphics[width = 9.0cm, angle = 0]{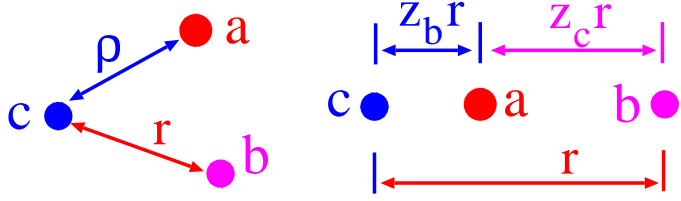}
\caption{
(Color online)
The color dipole structure (lhs) of the generic 3-parton state and 
(rhs) of the 3-parton state entering the color-dipole description
of fragmentation $a\to bc$ with 
formation of the $bc$ dipole of size $\br$.}
\label{fig:ABCdipole}
\end{center}
\end{figure}

If $\bb_a=\bb$ is the impact parameter of the projectile $a$, then
\beq
\bb_{b}=\bb+z_c\br, \quad\quad \bb_{c}=\bb-z_b\br\, ,
\label{eq:2.2}
\eeq
see Fig.~\ref{fig:ABCdipole}. 
Here $z_{b,c}$ stand for the fraction the lightcone momentum of the
projectile $a$ carried by the partons $b$ and $c$. 
The virtuality of the
incident parton $a$ is given by $Q_a^2=\bk_a^2$, where $\bk_a$ is the
transverse momentum of the parton $a$ in the incident proton 
(Fig.~\ref{fig:SingleJet-Rapidity}). For the
sake of simplicity we take the collision axis along the momentum of
parton $a$. The transformation between the transverse momenta in
the $a$-target and $p$-target reference frames is trivial,
$\bp_{b,c}^{(a)}=\bp_{b,c}^{(p)}+ z_{c,b}\bk_a$, below we cite all
the spectra in the $a$-target collision frame. We shall speak of the 
produced parton - or jet originating from this parton -
as the {\it leading} one if it carries a fraction of
the beam lightcone momentum $z\to 1$ and a {\it slow} one if it
is produced with $z\ll 1$. We speak of the produced parton as
a {\it hard} one if it is produced with large transverse momentum and
a {\it soft} one if its transverse momentum is small compared to
the so-called nuclear saturation scale to be defined below. 

\begin{figure}[!t]
\begin{center}
\includegraphics[width = 7.0cm,height=16cm, angle = 270]{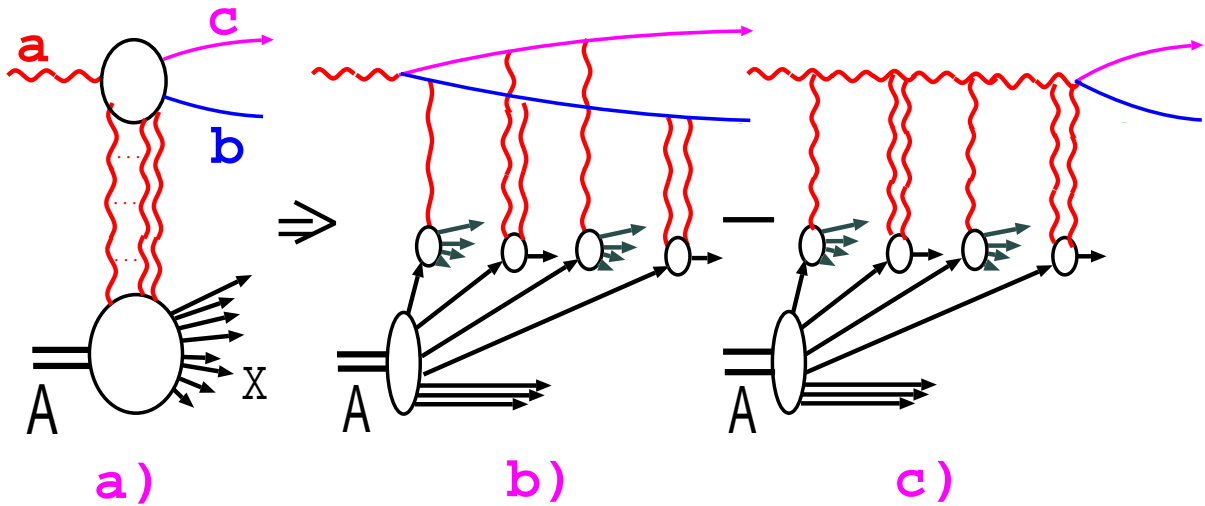}
\caption{
(Color online)
Typical contribution to the excitation amplitude for $g A \to q \bar{q} X$,
with multiple color excitations of the nucleus. 
{The amplitude receives contributions from processes that involve interactions 
with the nucleus after and before the virtual 
decay which interfere destructively.}}
\label{fig:single-jet_a_to_bc}
\end{center}
\end{figure}

By the conservation of
impact parameters, the action of the $S$-matrix on $\ket{a}_{phys}$
takes a simple form
\bea
&& \textsf{S}\ket{a}_{phys} =
S_a(\bb) \ket{a}_0 +
S_b(\bb_b) S_c(\bb_c)\Psi(z_b,\br) \ket{bc}_0 \nonumber \\
&&=S_a(\bb) \ket{a}_{phys} +  [ S_b(\bb_b) S_c(\bb_c) - S_a(\bb) ]
\Psi(z_b,\br) \ket{bc} \, . 
\label{eq:2.3} \eea 
In the last line we
explicitly decomposed the final state into the elastically
scattered $\ket{a}_{phys}$ and the excited state $\ket{bc}$.
The two terms in the latter describe a scattering on the target of
the $bc$ system formed way in front of the target and the
transition $a\to bc$ after the interaction of the state
$\ket{a}_0$ with the target, as illustrated in fig. \ref{fig:single-jet_a_to_bc}.
 The contribution from transitions
$a\to bc$ inside the target nucleus vanishes in the high-energy
limit 
\beq
x= 
{Q_a^2+M_{bc}^2\over W^2 x_B}\lsim x_A \,
\label{eq:2.3*}
\eeq
where $M_{bc}$ is the invariant mass of the excited $bc$ system,
$Q_a^2$ is the virtuality of the incident parton $a$, $W$ is the
total energy in the beam-target nucleon collision center of mass frame 
and $x_B$ is the fraction of the beam energy carried by the 
beam-parton $a$.

\begin{figure}[!t]
\begin{center}
\includegraphics[width = 5.5cm,angle=270]{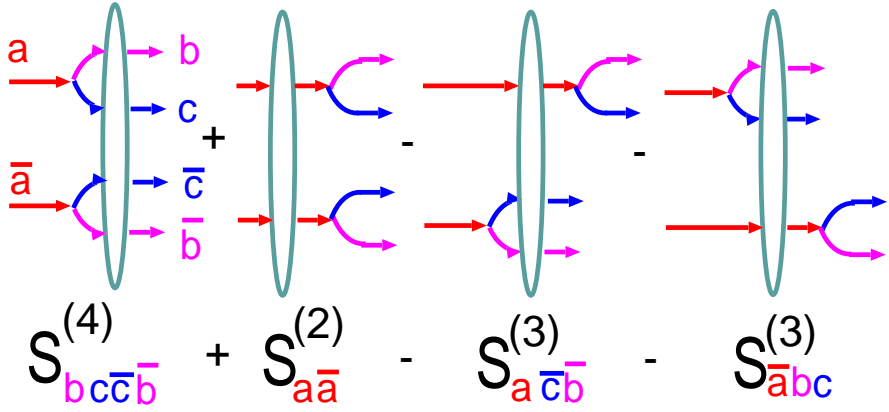}
\caption{
(Color online)
The $\textsf{S}$-matrix structure of the two-body density
matrix for excitation $a\to bc$.}
\label{fig:SingleJetDensityMatrix}
\end{center}
\end{figure}

The probability amplitude for the two-jet spectrum is given by the
Fourier transform \beq \int d^2\bb_b d^2\bb_c \exp[-i(\bp_b\bb_b +
\bp_c\bb_c)][ S_b(\bb_b) S_c(\bb_c) - S_a(\bb) ] \Psi(z_b,\br)
\label{eq:2.4} \eeq 
The differential cross section is proportional
to the modulus squared of (\ref{eq:2.4}) and one encounters the
contributions containing
\bea
S^{(2)}_{\bar{a}a}(\bb',\bb)&=& S_a^{\dagger}(\bb')S_a(\bb)
\nonumber \\ S^{(3)}_{\bar{a}bc}(\bb',\bb_b,\bb_c) &=&
S_a^{\dagger}(\bb')S_b(\bb_b) S_c(\bb_c), \nonumber \\
S^{(3)}_{\bar{b}\bar{c}a}(\bb,\bb_b',\bb_c') &=&
S_b^{\dagger}(\bb_b')S_c^{\dagger}(\bb_c') S_a(\bb) \nonumber \\
S^{(4)}_{\bar{b}\bar{c} c b}(\bb_b',\bb_c',\bb_b,\bb_c) &=&
S_b^{\dagger}(\bb_b')S_c^{\dagger}(\bb_c') S_c(\bb_c)S_b(\bb_b) \,
.
\label{eq:2.3**}
\eea
Here we suppressed the matrix elements of $S^{(n)}$ over the
target nucleon, for details see \cite{Nonlinear}. In the
calculation of the inclusive cross sections one averages over
the color states of the beam parton $a$, sums over color
states of final state partons $b,c$ and takes the matrix products of
$S^{\dagger}$ and $S$ with respect to the relevant color indices
entering $S^{(n)}$. Some of the technicalities of the derivation
of $S^{(n)}$ are presented in the Appendix A, here we cite
the resulting two-jet cross section: 
\bea
{d \sigma (a^* \to b(\bp_b) c(\bp_c)) \over dz d^2\bp_b d^2\bp_c } = {1 \over (2 \pi)^4} \int
d^2\bb_b d^2\bb_c d^2\bb'_b
 d^2\bb'_c ~~~~~~~~~~~~~~~~~~~~~~~~~~~~~~~~~~~~~~~~~~~\nonumber \\
\times \exp[i \bp_b
(\bb_b -\bb'_b) + i
\bp_c(\bb_c
-\bb_c')] \Psi(z,\bb_b -
\bb_c) \Psi^*(z,\bb'_b-
\bb'_c)~~~~~~~~~~~~~~~~~~  \nonumber \\
\{
S^{(4)}_{\bar{b}\bar{c} c b}(\bb_b',\bb_c',\bb_b,\bb_c) 
+ S^{(2)}_{\bar{a}a}(\bb',\bb) -
S^{(3)}_{\bar{b}\bar{c}a}(\bb,\bb_b',\bb_c')
- S^{(3)}_{\bar{a}bc}(\bb',\bb_b,\bb_c) \}
\label{eq:2.5} 
\eea 
The crucial point is that the hermitian conjugate
$S^{\dagger}$ can be viewed as the $S$-matrix for an antiparton
\cite{SlavaPositronium,NPZcharm,Nonlinear}. As a result,
$S^{(2)}_{\bar{a}a}(\bb',\bb)$ is an $S$-matrix for interaction
with the target of the $a\bar{a}$ state in which the antiparton
$\bar{a}$ propagates at the impact parameter $\bb'$. The averaging
over the color states of the beam parton $a$ amounts to taking the
color singlet $a\bar{a}$ state. The $S^{(3)}_{\bar{a}bc}$ and
$S^{(4)}_{\bar{b}\bar{c}cb}$ describe interaction of the
color-singlet $\bar{a}bc$ and $\bar{b}\bar{c}cb$ states,
respectively. This is shown schematically in 
Fig.~\ref{fig:SingleJetDensityMatrix}.

The $S^{(2)}$
and $S^{(3)}$ are readily calculated in terms of the 2-parton and
3-parton dipole cross sections \cite{NZ91,NZ94,NPZcharm}, general
rules for multiple scattering theory  
calculation of the coupled-channel $S^{(4)}$ are found
in \cite{Nonlinear} and need not be repeated here. 


The results for the two-jet cross sections will be
presented elsewhere, here we focus on the single-jet
problem. Integration over the transverse momentum $\bp_c$ of the jet $c$
gives
\beq
\bb_c =\bb_c' \,,
\label{eq:2.6*}
\eeq
\beq \bb - \bb' =
z_b(\br-\br')\, ,\quad \bb_b-\bb_b'=\br-\br'\, , \quad
\bb'-\bb_c=z_b\br'\, . 
\label{eq:2.6} 
\eeq 

In the generic case the 3-parton state $\bar{a}bc$ has the dipole
structure shown in Fig.~\ref{fig:ABCdipole} and the dipole cross section
$\sigma_{3\bar{a}bc}(\brho,\br)$. In the considered problem of the
$a\to bc$ excitation $\brho = z_b\br'$. The case of
$S^{(4)}_{\bar{b}\bar{c}cb}$ deserves special scrutiny. In the
general case, $S^{(4)}$ is a multichannel operator, an example is found in
\cite{Nonlinear}. Because in the single-particle spectrum 
$\bb_c' =\bb_c$, the unitarity relation
\beq
S_c^{\dagger}(\bb_c) S_c(\bb_c)=1
\label{eq:2.6**}
\eeq
leads to a fundamental simplification of $S^{(4)}_{\bar{b}\bar{c} c b} (\bb_b',\bb_c',\bb_b,\bb_c)$.
Specifically, in a somewhat symbolic form
\bea
S^{(4)}_{\bar{b}\bar{c} c b} (\bb_b',\bb_c',\bb_b,\bb_c)=
S_b^{\dagger}(\bb_b')S_c^{\dagger}(\bb_c) S_c(\bb_c)S_b(\bb_b)
=
S_b^{\dagger}(\bb_b')S_b(\bb_b)=S^{(2)}_{\bar{b}b}(\bb_b',\bb_b) \
, 
\label{eq:2.7}
\eea
i.e., the effect of interactions of the spectator parton $c$ vanishes 
upon the summation over all its color states and integration over all its transverse
momenta \cite{NPZcharm}. The coupled-channel operator
$S^{(4)}_{\bar{b}\bar{c} c b}$ then takes a single-channel form, some technical
points behind this derivation are clarified in Appendix A.  
The only trace of the observed parton $b$ having been produced in the fragmentation
$a\to bc$ is in the density matrix $\Psi(z_b,\bb_b -
\bb_c) \Psi^*(z_b,\bb'_b-\bb'_c)$ which defines the transverse momentum distribution 
of the parton $b$ in the beam parton $a$ and the partition $z_b \,,\,z_c= (1-z_b)$, of
the longitudinal momentum between the final state partons.
It will be more convenient to regard $S^{(2)}_{\bar{b}b}(\bb_b',\bb_b)$  as a function
of the transverse size of the $b\bar{b}$ dipole 
$\bb_b'-\bb_b$, and the impact parameter $\bB$ of the dipole--nucleus interaction,
and we shall from now on write
$S^{(2)}_{\bar{b}b}(\bb_b',\bb_b) \to S^{(2)}_{\bar{b}b}(\bB,\bb_b'-\bb_b)$.

The observation (\ref{eq:2.7}) can be generalized further to the cancellations
of beam spectator interactions. The detailed discussion
will be reported elsewhere, here we just quote the simple example
of open heavy flavour excitation. The above expounded formalism for the
fragmentation $g \to Q\bar{Q}$ can readily be generalized to
excitation of heavy flavour from the incident quark $q^* \to q'Q\bar{Q}$
which proceeds via $q^* \to q' g, ~~g\to Q\bar{Q}$.
The simplest case of direct importance for mid-rapidity heavy flavour production
in $pp,pA$ collisions is  when the produced $Q\bar{Q}$ pair carries a small 
fraction of the incident quark's momentum, i.e., the underlying
partonic subprocess is a collision of a slow gluon from the beam with a 
slow gluon from the target, $gg \to Q\bar{Q}$. The generalization of
eqs.~ (\ref{eq:2.6*}), (\ref{eq:2.6}) to this case shows that 
the beam quark $q^*$ and the spectator quark $q'$ would propagate at the same impact
parameter $\bb_q$. One readily finds that the amplitude of the  $q^* \to q'Q\bar{Q}$
will be given by eq.~(\ref{eq:2.3}) times the extra factor 
of $S_q(\bb_q)$. Upon the integration
over the transverse momentum of the spectator quark one gets
the equality of the corresponding impact parameters in the multiparton
$S$-matrix and its hermitian conjugate, eq.~(\ref{eq:2.6*}), and
upon the application of the unitarity relation (\ref{eq:2.6**}) the
effect of spectator interactions cancels out in complete analogy to
eq.~(\ref{eq:2.7}). The resulting $S^{(2)}_{\bar{b}b}(\bB,\bb_b'-\bb_b)$
only depends on the dipole parameter $\bb_b'-\bb_b$ and the overall
impact parameter, $\bB$, of the multiparton system. What will be left
of the beam quark $q^*$ is the wave function of its $q'g$ Fock state
which defines the longitudinal and transverse momentum density
in the beam of the gluon $g^*$ which 
excites into the $Q\bar{Q}$ state. The evident assumption
behind the spectator interaction cancellations is that the coherency
condition is fulfilled for all spectator partons; because the spectators
are faster than the observed parton, this 
weak condition does not impose any stringent constraints beyond
(\ref{eq:2.3*}).


\section{Color dipole master formula for free nucleons and nuclei}
\label{Sec:3}

For the free nucleon target
integration over the impact parameter $\bB$ gives the corresponding total
cross sections, for instance,
\bea
2\int d^2\bB \left[1- S^{(2)}_{\bar{b}b}(\bB,\bb_b'-\bb_b)\right]&=&
\sigma_{2,\bar{b}b}(\br-\br')\nonumber\\
2\int d^2\bB \left[1- S^{(3)}_{\bar{a}bc}(\bB,\bb' -\bb_c,\bb_b -\bb_c)\right]&=&
\sigma_{3,\bar{a}bc}(z_b\br',\br)\,.
\label{eq:3.1}
\eea
Then the single-jet spectrum for the free nucleon target can be cast in the form
\bea
{d \sigma_N(a^* \to b(\bp_b) c) \over dz_b d^2\bp_b} = {1 \over 2
(2 \pi)^2} \int d^2\br
 d^2\br' \exp[i \bp_b
(\br -\br')]
\Psi(z_b,\br) \Psi^*(z_b,\br') ~~~ \nonumber \\
\times \{\sigma_{3,\bar{a}bc} (z_b\br',\br)
+ \sigma_{3,a\bar{b}\bar{c}} ( z_b\br,\br')
- \sigma_{2,b\bar{b}}(\br
-\br')
-\sigma_{2,a\bar{a}}(z_b(\br
-\br')) \}
\label{eq:3.2}
\eea
In the generic $\bar{a}bc$ system there could be more
than one scheme of coupling to a color singlet state, in which case
$S^{(3)}_{\bar{a}bc}$ would be a coupled-channel operator, but
that is not the case with gluons and quarks. The coupled-channel
operator $S^{(4)}_{\bar{b}\bar{c} c b}$ also takes the
single-channel form (\ref{eq:2.7}).
As usual, we apply multiple scattering theory treating 
the nucleus as a dilute gas of color-singlet
nucleons, and upon the summation over
all the nuclear final states and application of the closure relation
the nuclear matrix elements of $S^{(n)}$ take the familiar
Glauber-Gribov form \cite{Glauber,Gribov}. 
The resulting nuclear single-jet spectrum per
unit area in the impact-parameter plane equals
\bea &&{d \sigma_A (a^* \to b(\bp_b) c)\over dz d^2\bp_b
d^2\bb} = {1 \over (2 \pi)^2} \int d^2\br
 d^2\br' \exp[i \bp_b
(\br -\br')]
\Psi(z_b,\br) \Psi^*(z_b,\br') ~~~ \nonumber \\
&&\times \left\{\Gamma_A[\bb,\sigma_{3,\bar{a}bc}(z_b\br',\br)]
+\Gamma_A[\bb,\sigma_{3,a\bar{b}\bar{c}} ( z_b\br,\br')]\right.
\nonumber\\ &&
\left.- \Gamma_A[\bb,\sigma_{2,b\bar{b}}(\br -\br')]
-\Gamma_A[\bb,\sigma_{2,a\bar{a}}(z_b(\br -\br'))] \right\} \, .
\label{eq:3.3} 
\eea 
We can use the Glauber-Gribov multiple scattering theory result
\bea 
\Gamma_A[\bb,\sigma]&=& 1 - \exp[-{1\over 2}\sigma T(\bb)]\, , 
\label{eq:3.4} 
\eea where
$T(\bb)= \ds \int_{-\infty}^\infty dr_z \, n_A(\bb,r_z)$ is the optical thickness
of the nucleus, and the nuclear density $n_A(\bb,r_z)$ is
normalized according to $\ds \int d^3\vec{r} \, n_A(\bb,r_z) = \int
d^2\bb\,  T(\bb) = A$.

Within the color-dipole formalism the $x$-dependence of 
the DIS structure function is driven by the $x$-dependence of the 
color-dipole cross section $\sigma_{q\bar{q}}(x,\br)$ which is governed
by the color-dipole form of the Balitskii--Fadin--Kuraev--Lipatov
(BFKL) equation \cite{NZZBFKL,NZ94,MuellerPatel}.
For  $x\gsim x_A$ DIS off a nucleus amounts to the sum of
incoherent scatterings off bound nucleons. The onset of 
nuclear coherence effects at $x\lsim x_A$ must be treated
with the color-dipole cross section $\sigma_{q\bar{q}}(x_A,\br)$ which has
been BFKL-evolved down to $x=x_A$. We recall that
for heavy nuclei well evolved shadowing sets in at $x_A\approx 10^{-2}$.
The considered lowest order 
excitation processes $a\to bc$  with rapidities $\eta_{b,c}\gsim \eta_A$
such that no more secondary partons are produced in the rapidity span 
between $\eta_b$ and $\eta_c$ or between $\eta_{b,c}$ and $\eta_A$ 
must also be treated in terms of $\sigma_{q\bar{q}}(x_A,\br)$. 
The values of $x_{b,c}$ attainable at RHIC are such
that jet production in the proton hemisphere of $pA$ collisions at
RHIC is dominated by precisely these lowest-order processes.
The allowance for higher-order processes
would amount to the $x$-evolution of the $k_{\perp}$-factorization
breaking and will be addressed to elsewhere. 


\section{$k_{\perp}$-factorization for open charm production off free nucleons}
\label{Sec:4}

Here the underlying pQCD excitation is $g \to Q\bar{Q}$,
i.e., $a=g,b=Q,c=\overline{Q}$. The process is symmetric under 
$Q \leftrightarrow \bar{Q}$ and for the sake of brevity we
we put $z\equiv z_Q$ and $\bp \equiv \bp_Q$. The relevant 2-parton and 
3-parton dipole cross sections
equal \cite{NZ94,NPZcharm} 
\bea
\sigma_3(x,z\br',\br) &=& {C_A \over 2 C_F} \Big[\sigma_{q\bar{q}}(x,z\br'-\br) 
+\sigma_{q\bar{q}}(x,z\br')
-\sigma_{q\bar{q}}(x,\br) \Big] + \sigma_{q\bar{q}}(x,\br) \, , \nonumber\\
\sigma_{gg}(x,z(\br-\br'))& = &{C_A \over C_F} \,
\sigma_{q\bar{q}}(x,z(\br-\br')) \, , \label{eq:4.1} \eea
where $C_A = N_c \,, \, C_F =\ds {N_c^2-1 \over 2 N_c}$ are the familiar 
quadratic Casimirs, and \cite{NZglue}
\bea \sigma_{q\bar{q}}(x,\br) &=& \int d^2\bkappa f (x,\bkappa)
[1-\exp(i\bkappa\br)]\,,\nonumber \\
f (x,\bkappa)&=& {4\pi \alpha_S(r)\over N_c}\cdot {1\over \kappa^4} \cdot {\cal
F}(x,\kappa^2)\, . \label{eq:4.2} \eea
$ \sigma_{q\bar{q}}(x,\br)$ 
is the dipole cross section for the $q\bar{q}$ color dipole and
\beq
{\cal F}(x,\kappa^2) = {\partial G(x, \kappa^2) \over \partial \log \kappa^2}
 \label{eq:4.2*}
\eeq is  the unintegrated gluon density in the
target nucleon (our normalization of $f(x_A,\bkappa)$ and of the related
Weizs\"acker-Williams gluon density $\phi(\bb,x_A,\bkappa)\equiv
\phi_{WW}(\bb,x_A,\bkappa)$, see below, are slightly different from
the ones used in \cite{Nonlinear}). In the momentum space calculations
the running coupling $\alpha_S(r)$ must be taken at an appropriate
hard scale. 

Applying the technique
developed in \cite{Nonlinear}, upon the relevant Fourier
transformations one readily finds
\bea &&{2 (2\pi)^2 d\sigma_N(g^*
\to Q\bar{Q}) \over dz d^2\bp} = 
\int d^2\bkappa f(x_A,\bkappa)\nonumber\\
&&\Big\{ {N_c^2 \over N_c^2-1}\Big( |\Psi(z,\bp)
-\Psi(z,\bp+z\bkappa)|^2 + |\Psi(z,\bp+\bkappa)
-\Psi(z,\bp+z\bkappa)|^2\Big) \nonumber \\
&& + {1 \over N_c^2-1} |\Psi(z,\bp) - \Psi(z,\bp+\bkappa)|^2 \Big\}
\label{eq:4.3*} \\
&& = \int d^2\bkappa f(x_A,\bkappa)
\Biggl({C_A\over
2C_F}\Big\{ |\Psi(z,\bp) -\Psi(z,\bp+z\bkappa)|^2 + |\Psi(z,\bp+\bkappa)
-\Psi(z,\bp+z\bkappa)|^2 \nonumber\\
&& -|\Psi(z,\bp) -\Psi(z,\bp+\bkappa)|^2 \Big\} 
 + |\Psi(z,\bp) - \Psi(z,\bp+\bkappa)|^2 \Biggl) \, . \label{eq:4.3}
\eea
Here we show explicitly that the target gluon density $f(x_A,\bkappa)$
enters with the boundary condition argument $x_A$. 
The expression  $|\Psi(z,\bp) -\Psi(z,\bp+z\bkappa)|^2$ for the
$Q\bar{Q}$ Fock state of the gluon is obtained from its counterpart
for the transverse photon cited in \cite{Nonlinear} by the substitutions
$Q^2\to Q_a^2$ and $N_c\alpha_{em}e_Q^2 \to T_F\alpha_S$,
see Appendix B. 

Here we emphasize that Eqns.~(\ref{eq:4.3*},\ref{eq:4.3}) account
fully for the transverse momenta of all involved partons, no
further smearing of the obtained spectra for the intrinsic 
transverse momentum of partons is needed, furthermore, it would be
illegitimate.
{Early discussions of charm production in hadron--nucleon collisions 
in the framework of $k_\perp$--factorization can be found in  \cite{Collins:1991ty},
for reviews and a guide to recent literature we refer to 
\cite{Andersson:2002cf}. The rather straightforward representation 
(\ref{eq:4.3*},\ref{eq:4.3})
of the heavy quark cross section in terms of light--cone wavefunctions 
for the transition $g^* \to Q\bar{Q}$ is new, as are the related 
expressions for quark/gluon production to be found below.} 

Some comments on
this simple formula are in order. First, although we spoke of the
charm production, as soon as the transverse momentum $\bp$ is
perturbatively large eq. (\ref{eq:4.3}) will fully be applicable to
light quark jets and there is an obvious connection between
(\ref{eq:4.3}) and the NLL real correction to the BFKL kernel from
light $q\bar{q}$ pair production, but dwelling into that is beyond
the scope of the present communication. Second, we presented our
result for two different groupings of terms. While in the first
grouping (\ref{eq:4.3*}) the result is given as a manifestly positive sum of
squares, in the second one (\ref{eq:4.3}) we decomposed the cross section into
'nonabelian' , $\propto C_A/2C_F$, and 'abelian' pieces. Putting
$C_A\to 0$ one would switch off the nonabelian coupling of gluons
to gluons and obtain indeed the same formula as for the
single-quark spectrum from the $\gamma^* \to q\bar{q}$ transition,
i.e., for inclusive DIS \cite{Nonlinear}. Third, note that the
same 'abelian' limit is obtained at the boundaries $z = 0$ and
$1-z=0$. In the both cases that can be traced to the fact either
the spectator quark for $z=0$, or the observed antiquark for $z=1$
would propagate at precisely the same impact parameter as the
parent gluon so that the non-abelian color structure of the
$gQ\overline{Q}$ is simply not resolved. Still the physics of the
two cases is different: in the limit of $z=0$ the `initial state'
$gg$ dipole interaction effectively drops out, whereas at $z=1$
there are subtle cancellations between the $gg$ dipole cross
section and the octet-octet dipole components of the 3-parton
cross sections. As we shall see below, the latter cancellations
will be upset for nuclear targets.
Fourth, in the large-$N_c$ limit $C_A=2C_F$ and for later 
reference we quote
\bea 
&&{2 (2\pi)^2 d\sigma_N(g^* \to Q\bar{Q}) \over dz
d^2\bp}\Big|_{N_c\gg 1} = \nonumber\\
&&=\int d^2\bkappa f(x_A,\bkappa) \,
\Big\{|\Psi(z,\bp) -\Psi(z,\bp+z\bkappa)|^2 + |\Psi(z,\bp+\bkappa) -
\Psi(z,\bp+z\bkappa)|^2 \Big\} \, .
\label{eq:4.4}
\eea

Based on the discussion of spectator effects in section \ref{Sec:3} and
anticipating the discussion in section \ref{Sec:6}, we note that the open-charm
production can be presented in a manifestly beam-target symmetric 
form. One can multiply the above derived cross section
$d\sigma_Q^{(g^*\to Q\bar{Q})}(z,x_A,\bp)$ 
by the unintegrated flux of gluons of transverse momentum $\bk_a$ 
and lightcone momentum fraction $x_B$
in the beam proton, ${\cal F}(x_B,\bk_a)$, take the wave functions 
of Appendix B at $Q^2=\bk_a^2$, and calculate the charm spectrum for
$pp,pA$ collisions as
\beq
 d\sigma_{N,A} (z,x_A,\bp) = {1\over x_B}\int d^2\bk_a {\cal F}(x_B,\bk_a) 
d\sigma_{N,A}^{(g^*\to Q\bar{Q})}(z,x_A,\bp+z\bk_a)\,.
\label{eq:4.5}
\eeq
The usual kinematical constraint, 
$ \ds x_B x_A W^2 = {\bp^2+m_Q^2 \over z(1-z)} \, ,
$
is understood. The fully inclusive single-parton cross section is
obtained from (\ref{eq:4.5}) upon integration over $x_B$ keeping 
fixed the longitudinal momentum of the observed parton $zx_B$. 
We stick to $z$ because (i) the transverse momentum spectra 
 exhibit a highly nontrivial dependence on $z$ and (ii) 
understanding the $z$-dependence is crucial for the LPM effect. 

\begin{figure}[!t]
\begin{center}
\includegraphics[width = 6.0cm]{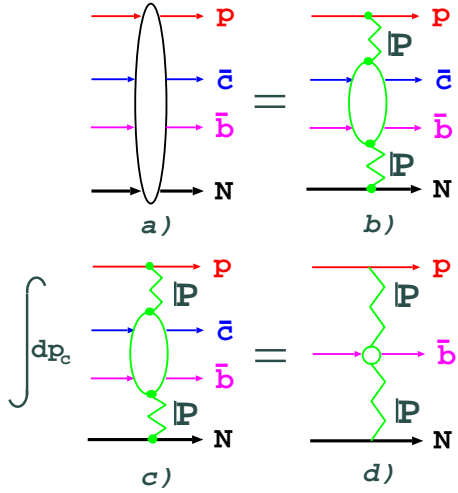}
\caption{(Color online)
(a) The Kancheli-Mueller diagram for the two-parton
inclusive cross section in proton-proton collisions, (b) its 
representation in terms of the pomeron exchange and (c) transition
to the optical theorem for the single-parton
inclusive cross section (d) upon the integration over
the phase space of the parton $c$. } 
\label{fig:SingleJetKancheli_N}
\end{center}
\end{figure}

Within the $k_{\perp}$-factorization the microscopic QCD description
of the pomeron exchange is furnished by the  unintegrated gluon 
densities. To this end one can look at the open-charm cross section
(\ref{eq:4.5}) through the prism of the Kancheli-Mueller diagrams
\cite{KancheliMueller} for inclusive spectra as depicted 
in a somewhat symbolic way in figure (\ref{fig:SingleJetKancheli_N}).  
The $\bar{c}\bar{b}\Pom \bar{c}\bar{b}\Pom$ vertex which enters the
pomeron-exchange representation (b) for the generic Kancheli-Mueller
discontinuity of the 4-body forward scattering amplitude is a highly
nonlocal one, its properties will be discussed to more detail elsewhere.
Upon integration over the momentum of the parton $c$ one obtains the 
Kancheli-Mueller diagram (d) for the single-parton spectrum, for the
nonlocal properties of the $\bar{b}\Pom \bar{b}\Pom$ vertex in
the diagram (d) see Sect.\ref{Sec:6.4.1}.

\section{Nuclear case: nonlinear $\bk_\perp$--factorization}
\label{Sec:5}

The driving term of small-$x$ nuclear structure functions and single-quark
spectra from the $\gamma^* \to q\bar{q}$ excitation off a nucleus were shown
to take the familiar $k_{\perp}$-factorization form in terms of the
collective nuclear Weizs\"acker--Williams gluon density $\phi(\bb,x_A,\bkappa)$
per unit area in the impact parameter plane 
as defined in \cite{NSSdijet,NSSdijetJETPLett,Nonlinear,NonlinearJETPLett}:
\bea \Gamma_{A}[\bb,\sigma_{q\bar{q}}(x,\br)] \equiv \int d^2\bkappa
\phi(\bb,x,\bkappa) \Big[1 - \exp[i\bkappa\br] \Big] \, .\label{eq:5.1} \eea
It satisfies the sum rule
\beq
\int d^2\bkappa \phi(\bb,x,\bkappa) = 1 - \exp[-{1\over 2} \sigma_0(x)
T(\bb) ] = 1 - S_{abs}(\bb) \, .
\label{eq:5.2}
\eeq
where $\ds \sigma_{0}(x)=\int d^2 \bkappa f(x,\bkappa)$ is the 
dipole cross section for large dipoles. 
It is conveniently interpreted in the spirit of 
additive quark counting for soft hadronic processes as
$ \sigma_0(x) = \sigma_{qN}(x) + \sigma_{\bar{q}N}(x) = 2 \, \sigma^\Pom_{qN}(x)$,
so that the meaning of the factor
\beq
S_{abs}(\bb) =  \exp[-{1\over 2} \sigma_0(x)
T(\bb) ] =  \exp[- \sigma^\Pom_{qN}(x) T(\bb) ]
\eeq
as representing intranuclear absorption becomes evident.
 
\begin{figure}[!t]
\begin{center}
\includegraphics[width = 12.0cm]{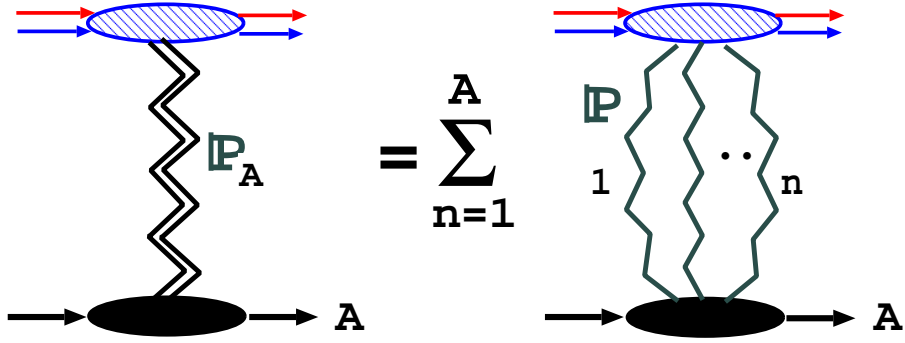}
\caption{
(Color online)
Exchange by a nuclear pomeron $\Pom_A$ in color dipole-nucleus
scattering as a sum of multiple two-gluon pomeron exchanges. } 
\label{fig:SingleJetNuclearPomeron}
\end{center}
\end{figure}

\begin{figure}[!t]
\begin{center}
\includegraphics[width = 14.0cm]{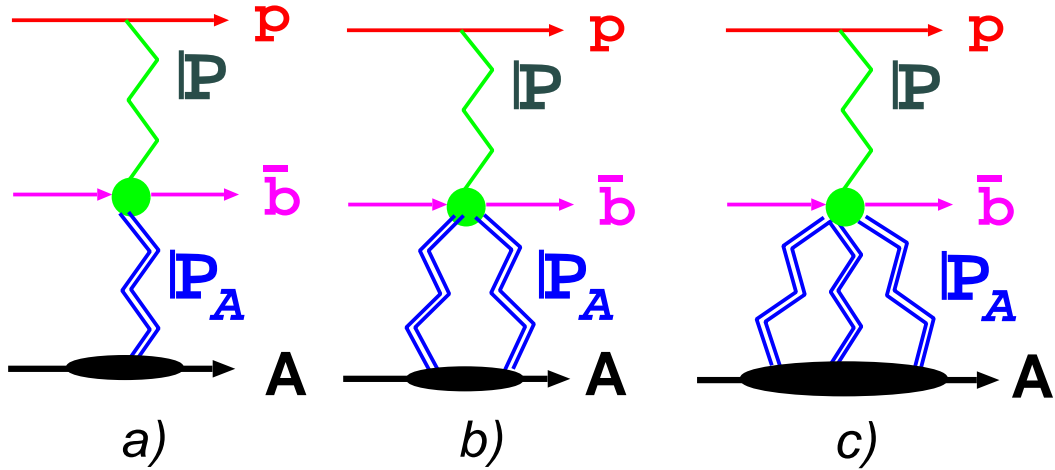}
\caption{
(Color online)
The Kancheli-Mueller diagrams for the mid-rapidity 
single-particle spectrum form $gg$ subcollisions in $pA$ collisions.
The double-zigzag lines describe an exchange by the nuclear
pomeron $\Pom_A$. The open-charm spectrum in $pp$ collisions is 
dominated by the diagram (a) with $\Pom_A \to \Pom$, 
see Fig.~\ref{fig:SingleJetKancheli_N}, the open
charm spectrum in central $pA$ collisions is dominated by
the diagram (b) and the gluon jet production is dominated 
by the diagram (c). } 
\label{fig:SingleJetKancheli_A}
\end{center}
\end{figure}

Eqs. (\ref{eq:3.3}), (\ref{eq:4.1}) and (\ref{eq:4.2}) allow a
straightforward calculation of the nuclear single-jet spectra.
Nice analytic forms of the nuclear $k_{\perp}$-factorization are 
obtained in the large-$N_c$ limit, the $1/ N_c^2$
corrections can readily be derived following the considerations
in \cite{Nonlinear} and will not be discussed here. The crucial
point about the large-$N_c$ approximation is that
\bea
\sigma_3(x,z\br',\br) \Longrightarrow \sigma_{q\bar{q}}(x,z\br'-\br) +\sigma_{q\bar{q}}(x,z\br')\, ,
\nonumber\\
\sigma_{gg}(x,z(\br-\br'))\Longrightarrow 2\sigma_{q\bar{q}}(x,z(\br-\br'))
\label{eq:5.3} \eea so that 
\bea 
\Gamma_A[\bb,\sigma_{3}(
z\br',\br)]&=& \Gamma_A[\bb,\sigma_{q\bar{q}}(x, z\br'-\br)]+
\Gamma_A[\bb,\sigma_{q\bar{q}}(x,z\br')] \nonumber\\
&-&\Gamma_A[\bb,\sigma_{q\bar{q}}(x, z\br'-\br)]\cdot
\Gamma_A[\bb,\sigma_{q\bar{q}}(x,z\br')]\, , \nonumber\\
\Gamma_A[\bb,\sigma_{gg}(x, z(\br-\br'))]&=& 2\Gamma_A[\bb,\sigma_{q\bar{q}}(x,
z(\br-\br'))]-\Gamma_A^2[\bb,\sigma_{q\bar{q}}(x, z(\br-\br'))]\,. \label{eq:5.4}
\eea 
Then after some algebra we obtain
\bea &&{(2\pi)^2 d\sigma_A(g^* \to Q\bar{Q}) \over dz d^2\bp d^2\bb}= \nonumber\\
&& S_{abs}(\bb) \cdot \int d^2\bkappa \phi(\bb,x_A,\bkappa) \Big\{|\Psi(z,\bp) -
\Psi(z,\bp+z\bkappa) |^2 + |\Psi(z,\bp +\bkappa) -
\Psi(z,\bp+z\bkappa)|^2\Big\} \nonumber \\
&&+ \int d^2\bkappa_1 d^2\bkappa_2 \phi(\bb,x_A,\bkappa_1) \phi(\bb,x_A,\bkappa_2)
|\Psi(z,\bp+\bkappa_2) -\Psi(z,\bp+z\bkappa_1+z\bkappa_2) |^2 \, .
\label{eq:5.5}
\eea
One can associate with $\phi(\bb,x_A,\bkappa_1)$ the exchange of a
nuclear pomeron $\Pom_A$ which as indicated in Fig.~\ref{fig:SingleJetNuclearPomeron}
sums multiple exchanges
of the pomeron $\Pom$ in interactions with nuclear targets
(\cite{Nonlinear,PionDijet}, see also Sect. \ref{Sec:6.2}). The 
first term is an exact counterpart of the free-nucleon 
result: in terms of the collective nuclear gluon
density $\phi(\bb,x_A,\bkappa)$ it has exactly the same linear
$k_{\perp}$-factorization form as the free nucleon result (\ref{eq:4.4})
in terms of $f(x_A,\bkappa)$. 
As such it corresponds to the Kancheli-Mueller diagram
of Fig.~\ref{fig:SingleJetKancheli_A}a. Despite the similarity
of diagrams of  Fig.~\ref{fig:SingleJetKancheli_A}a and
Fig.~\ref{fig:SingleJetKancheli_N}d, the two inclusive spectra
are different because $\Pom_A \neq \Pom$. Furthermore, this
contribution to the nuclear spectrum is suppressed by the absorption factor
$S_{abs}(\bb)$ and is entirely negligible for heavy nuclei. In this
case only the second term survives, which is a manifestly nonlinear
functional of the collective nuclear glue $\phi(\bb,x_A,\bkappa)$.
It can be associated with the Kancheli-Mueller diagram of  
Fig.~\ref{fig:SingleJetKancheli_A}b.
This completes our demonstration of the $k_{\perp}$-factorization 
breaking for single-jet spectra: the pattern of 
$k_{\perp}$-factorization changes
from the conventional linear at the periphery of a nucleus to
the nonlinear - quadratic - one for small impact parameters.
In retrospect, the finding of linear
$k_{\perp}$-factorization for the single quark jet spectrum in
$\gamma^*\to q\bar{q}$ excitation can be traced back to the beam
photon being colorless, by which the single-jet spectrum
becomes an abelian observable \cite{Nonlinear}. In contrast to
that, the origin of the nonlinear term in (\ref{eq:5.5}) is in the
nonabelian nature of the 3-parton cross section (\ref{eq:4.1}), (\ref{eq:5.3}).

The $z$-dependence of the jet spectra enters through the wave-functions.
To this end notice the strikingly different wave-function structure of the
free-nucleon cross section (\ref{eq:4.5}) and its nuclear counterpart
(\ref{eq:5.5}). This suggests a nontrivial dependence of the
$\bp$-spectra on the longitudinal variable $z$ which will be
different for the free-nucleon and nuclear targets. Vice versa, 
the $z$-distribution will vary with $\bp$ and the target mass number
which can be called the LPM effect for open charm production. 


\section{The Cronin effect for open charm}
\label{Sec:6}


\subsection{The charm $\bp$-distribution: nuclear vs. free nucleon targets}
\label{Sec:6.1}

The impact of nonlinear
$k_{\perp}$-factorization is quantified by the ratio of the free-nucleon 
and nuclear single-jet spectra,
$$
R_{A/N}(z;x_A,\bp)= {d\sigma_A\over A d\sigma_N}\,.
$$
This ratio exhibits both shadowing, $R_{A/N}(z;x_A,\bp) < 1$,
and antishadowing, $R_{A/N}(z;x_A,\bp) > 1$, properties, in the literature the
latter is usually referred to as the Cronin effect. Our discussion of the
Cronin effect is to a large extent based on
the ideas and technique of the pQCD derivation of nuclear properties of the
coherent diffractive breakup of pions into hard dijets  
\cite{NSSdijet,NSSdijetJETPLett}. 

The principal points are best seen for central impact 
parameters, i.e., for
$S_{abs}(\bb) \to 0$:
\bea
&&{(2\pi)^2 d\sigma_A(g^* \to Q\bar{Q}) \over dz d^2\bp d^2\bb}= \nonumber\\
&&\int d^2\bkappa_1 d^2\bkappa_2 \phi(\bb,x_A,\bkappa_1)
\phi(\bb,x_A,\bkappa_2) |\Psi(z,\bp+\bkappa_2)
-\Psi(z,\bp+z\bkappa_1+z\bkappa_2) |^2 \, .
\label{eq:6.1.1} 
\eea
In all the comparisons of the free-nucleon and nuclear target
observables we use the large-$N_c$ approximation without further notice.

The results simplify greatly in two interesting cases.
We start with the limit
of slow quarks, $z\ll 1$, when the free nucleon spectrum becomes 
\bea &&{2 (2\pi)^2 d\sigma_N(g^*
\to Q\bar{Q}) \over dz d^2\bp}\Biggr|_{z\ll 1} = \int d^2\bkappa f(x_A,\bkappa)
|\Psi(z,\bp) - \Psi(z,\bp+\bkappa)|^2  \, . 
\label{eq:6.1.2}
\eea
In the limit of $z\ll 1$ the entire dependence of the nuclear spectrum (\ref{eq:6.1.1})
on $\bkappa_1$ is confined to
the normalization integral $\ds \int d^2\bkappa_1
\,\phi(\bb,x_A,\bkappa_1)=1-S_{abs}(\bb)$ what
lowers the nonlinearity of the nuclear cross section. Then, in conjunction with the
linear term, $\propto S_{abs}(\bb)$, in (\ref{eq:5.5}), we find
a result applicable at all impact parameters:
 \bea 
{(2\pi)^2 d\sigma_A(g^* \to
Q\bar{Q}) \over dz d^2\bp d^2\bb}\Biggr|_{z\ll 1}= \int d^2\bkappa
\phi(\bb,x_A,\bkappa) |\Psi(z,\bp+\bkappa) -\Psi(z,\bp) |^2 \, .
\label{eq:6.1.3} 
\eea
We evidently recovered the linear $k_{\perp}$-factorization
result (\ref{eq:6.1.2}), now 
in terms of the nuclear collective glue $\phi(\bb,x_A,\bkappa)$.
The physics behind this finding is the above discussed abelianization
of $g\to Q\overline{Q}$ at $z\ll 1$, by which it becomes
equivalent to $\gamma^* \to Q\overline{Q}$. 

The second simple result is found for leading quarks
with $z\to 1$. Here the free-nucleon
cross section again takes the form (\ref{eq:6.1.2}), whereas
the nuclear single-jet spectrum can be cast in the convolution form 
\bea
{ d\sigma_A(g^* \to Q\bar{Q}) \over dz d^2\bp d^2\bb}\Biggr|_{z\to 1}=
\int d^2\bkappa \phi(\bb,x_A,\bkappa) {d\sigma_A(x,\bp -\bkappa)\over dz d^2\bp d^2\bb}\Biggr|_{z\ll 1} 
+S_{abs}(\bb)
{d\sigma_A(x,\bp)\over dz d^2\bp d^2\bb}\Biggr|_{z\ll 1}\, .
\label{eq:6.1.4}\, 
\eea
The leading term in (\ref{eq:6.1.4}) is
a manifestly nonlinear - quadratic - functional of $\phi(\bb,x_A,\bkappa)$. 
Evidently, such a convolution would enhance the nuclear
effects compared to the linear
$k_{\perp}$-factorization formula (\ref{eq:6.1.3}). Arguably, when
$z$ varies from $z=0$ to $1$, the $\bp$-dependence of the nuclear
charm spectra shall interpolate between the
ones given by equations (\ref{eq:6.1.3}) and (\ref{eq:6.1.4}).


\subsection{Properties of the collective nuclear gluon density}
\label{Sec:6.2}

The fundamental quantity of our formalism 
is the collective nuclear glue $\phi(\bb,x_A,\bkappa)$ per unit area in 
the impact parameter plane 
\cite{NSSdijet,NSSdijetJETPLett,Nonlinear}.
It is characterized by a new scale - the saturation scale, 
\beq
Q_A^2(\bb,x_A)\approx {4\pi^2 \over N_c}\alpha_S(Q_A^2)G(x_A,Q_A^2)T(\bb)\,, 
\label{eq:6.2.1}
\eeq
a related quantity appears in other approaches to nuclear
parton densities \cite{Mueller:1989st,CGC}.
At soft transverse momenta, $\bkappa^2 \lsim Q_A^2(\bb,x_A)$,
the nuclear collective glue 
exhibits a plateau which signals the saturation
of parton densities. A useful parameterization in the plateau
region is
\beq
\phi(\bb,x_A,\bkappa) \approx {1\over \pi}\cdot {Q_A^2(\bb,x_A)
\over (\bkappa^2 + Q_A^2(\bb,x_A))^2}\,. 
\label{eq:6.2.2}
\eeq 
For hard gluons, $\bkappa^2 \gsim Q_A^2(\bb,x_A)$, it is convenient
to cite the unintegrated gluon density per bound nucleon: 
\bea
f_A(\bb,x_A,\bkappa) &=& {\phi(\bb,x_A,\bkappa) \over {1\over 2}T(\bb)}\nonumber\\
&=& 
f(x_A,\bkappa )\left[1+ {\gamma^2 \over 2}\cdot 
{\alpha_S(\bkappa^2)G(x_A, \bkappa^2) \over \alpha_S(Q_A^2)G(x_A,Q_A^2)}
\cdot {Q_A^2(\bb,x_A) \over \bkappa^2}\right] \nonumber\\
&=& 
f(x_A,\bkappa )\left[1+ {2\pi^2\gamma^2 \over N_c\bkappa^2}\cdot 
\alpha_S(\bkappa^2)G(x_A, \bkappa^2) T(\bb)\right] \nonumber\\
&=&f(x_A,\bkappa )\left[1+ \Delta_{HT}(\bb,x_A,\bkappa)\right]  \,,
\label{eq:6.2.3}
\eea
where $\gamma \approx 2$ is an exponent of the large-$\bkappa^2$ tail $f(x_A,\bkappa)
\propto (\bkappa^2)^{-\gamma}$. Here we show explicitly the leading 
nuclear higher twist correction, $\Delta_{HT}(\bb,x_A,\bkappa)$, 
which gives rise to a nuclear antishadowing
effect \cite{NSSdijet,NSSdijetJETPLett,Nonlinear}. 
To this end, the saturated gluon density (\ref{eq:6.2.2}) can be regarded
as a resummation of all higher twist terms 
$\sim (Q_A^2(\bb,x_A)/\bkappa^2)^n$. 
Note the very strong nuclear suppression of the gluon density per bound nucleon
$f_A(\bb,x_A,\bkappa )$ in the soft plateau region, $\bkappa^2 \lsim Q_A^2(\bb,x_A)$:
\beq
f_A(\bb,x_A,\bkappa) \approx {2 \over \pi Q_A^2(\bb,x_A) T(\bb)} \propto 
{1 \over T^2(\bb)} \propto A^{-2/3}
\label{eq:6.2.4}
\eeq

The explicit expansion for $\phi(\bb,x_A,\bkappa)$ 
in terms of the collective
gluon density of $j$ overlapping nucleons in the Lorentz-contracted nucleus
\cite{Nonlinear,NSSdijet,NSSdijetJETPLett} is
\beq
\phi(\bb,x_A,\bkappa) =S_{abs}(\bb) \sum_{j=1}{1\over j!}\cdot 
\left({T(\bb)\over 2}\right)^j\cdot f^{(j)}(x_A,\bkappa)
\label{eq:6.2.5}
\eeq
where $f^{(j)}(x,\bkappa)$ is the $j$-fold convolution, 
\beq
f^{(j)}(x,\bkappa)= \int \delta(\sum_{i=1}^{j}\bkappa_i -\bkappa)
\prod_{i=1}^j f(x,\bkappa_i)d^2\bkappa_i \,.
\label{eq:6.2.6}
\eeq
Finally, we notice that $\phi(\bb,x_A,\bkappa)$ satisfies the important sum rule
(see also \cite{Kharzeev:2003wz,Iancu:2004bx})
\beq
\int d^2\bkappa\, \bkappa^2 \,
\left[\phi(\bb,x_A,\bkappa)-{1\over 2}T(\bb)f(x_A,\bkappa)\right]
=0\, ,
\label{eq:6.2.7}
\eeq
which readily follows from the property that $\Gamma_{A}[\bb,\sigma(x,\br)]-
{1\over 2}T(\bb)\sigma(x,\br)$ vanishes $\propto \sigma^2(x,\br)$ as $\br^2 \to 0$.
The effect of nuclear suppression for nuclear soft gluons, see eqs. (\ref{eq:6.2.2}) and
(\ref{eq:6.2.4}), is compensated for by nuclear antishadowing (\ref{eq:6.2.3}) 
for hard glue. 

The familiar integrated gluon density
equals
\beq
G(x,Q^2) = {N_c \over 4\pi^2 \alpha_S} \int^{Q^2} d^{2}\bkappa\, \bkappa^2 \,f(x,\bkappa)\,
\label{eq:6.2.8}
\eeq
its nuclear counterpart per bound nucleon, $G_A(\bb,x_A,\bkappa^2)$, can be defined
likewise in terms of $f_A(\bb,x_A,\bkappa)$. Then the corollary of the
sum rule (\ref{eq:6.2.7}) is that for $Q^2 \gg Q_A^2(\bb,x_A)$
\bea
&&G_A(\bb,x_A,Q^2) -G(\bb,x_A,Q^2) = -\int_{Q^2} d^2\bkappa\, \bkappa^2 \,
f(x_A,\bkappa)\Delta_{HT}(\bb,x_A,\bkappa)
  \nonumber\\
&\approx&-{2\gamma^2 \pi^2 \alpha_S(Q^2) \over N_c}\cdot {\alpha_S(Q^2)G(x_A,Q^2) \over
\alpha_S(Q_A^2(\bb,x_A))G(x_A,Q_A^2(\bb,x_A)) }\cdot {Q_A^2(\bb,x_A)\over Q^2}\cdot
{\partial G(x_A,Q^2) \over \partial \log Q^2} \nonumber\\
&\approx &
-{\gamma^2 \pi^2 \alpha_S(Q^2) \over N_c}\cdot {\alpha_S(Q^2) Q_A^2(\bb,x_A)\over
\alpha_S(Q_A^2(\bb,x_A))G(x_A,Q_A^2(\bb,x_A))} \cdot 
{\partial G^2(x_A,Q^2) \over \partial  Q^2} 
\,,
\label{eq:6.2.9}
\eea
where we  made use of the expansion
(\ref{eq:6.2.3}).


\subsection{The Cronin effect: nuclear shadowing for $\bp^2 \lsim
Q_A^2(\bb,x_A)$}
\label{Sec:6.3}

For the sake of illustration we start with the case of $z\ll 1$ and assume that
the saturation scale is so large that even for $m_Q^2 \ll \bp^2 \ll Q_A^2(\bb,x_A)$
the jets are still perturbatively hard. Because the dominant
contribution to (\ref{eq:6.1.3}) comes from $\bkappa^2 \sim Q_A^2(\bb,x_A)$,
the expansion (\ref{eq:B.3}) of Appendix B simplifies
further, and the $\bkappa$ integration can be carried out explicitly
\bea 
{(2\pi)^2 d\sigma_A(g^* \to
Q\bar{Q}) \over dz d^2\bp d^2\bb}\Biggr|_{z\ll 1}&=& 
2T_F\alpha_S\cdot {1\over \bp^2}\int d^2\bkappa
\phi(\bb,x_A,\bkappa)
\nonumber\\ 
&=& 
[1-S_{abs}(\bb)]\cdot2T_F\alpha_S \cdot {1\over \bp^2}\,.
\label{eq:6.3.2} 
\eea
It is entirely dominated by the contribution from $\bkappa^2 \gsim \bp^2$, which
in the language of evolution amounts to the anti-DGLAP splitting 
of collective nuclear gluons with
large transverse momentum $\bkappa$ into quarks with small transverse
momentum $\bp$. 

The integration over all impact parameters gives 
\beq
2\int d^2\bb [1-S_{abs}(\bb)]= 2 \int d^2\bb \left\{1-\exp\left[-{1\over 2}\sigma_0(x_A) T(\bb)
\right]\right\}=\sigma_A[\sigma_0(x_A)]\,,
\label{eq:6.3.3} 
\eeq
which is the familiar Glauber-Gribov nuclear total
cross section $\sigma_A[\sigma_0]$ for a projectile with 
the free-nucleon cross section $\sigma_0$, and 
\beq
R_{A/N}(z;x_A,\bp) = {\sigma_A[\sigma_0(x_A)] \over A\sigma_0(x_A)} \sim A^{-1/3} < 1
\label{eq:6.3.4} 
\eeq
A marginal caveat is that for the fixed quark-jet momentum the 
inequality $\bp^2 \ll Q_A^2(\bb,x_A)$ does not hold uniformly 
over all impact parameters $\bb$, it breaks down for the most 
peripheral interactions, the peripheral contribution does not 
affect the shadowing property $R_{A/N}(z;x_A,\bp)< 1$, though. 

Now turn to the opposite limit of $z\to 1$. Here the free-nucleon cross
section is the same as for $z\ll 1$. The convolution representation 
(\ref{eq:6.1.4}) for the nuclear spectrum 
is already suggestive of stronger nuclear effects. 
The analysis by the straightforward use of (\ref{eq:6.1.1}) 
in which  one can neglect 
$\bp$ in the arguments of the wave functions is still simpler:
\bea
&&{(2\pi)^2 d\sigma_A(g^* \to Q\bar{Q}) \over dz d^2\bp d^2\bb}\Biggr|_{z\to 1}= 
\nonumber\\
&&2T_F\alpha_S\int d^2\bkappa_1 d^2\bkappa_2 \phi(\bb,x_A,\bkappa_1)
\phi(\bb,x_A,\bkappa_2) {\bkappa_1^2 \over (\bkappa_2^2+m_Q^2)( (\bkappa_1+\bkappa_2)^2+m_Q^2)} .
\label{eq:6.3.5} 
\eea
The integral in the r.h.s. will be dominated by the two logarithmic contributions
from (i) $\bkappa_2^2 \ll \bkappa_1^2$ and (ii) $(\bkappa_2+ \bkappa_1)^2 \ll \bkappa_1^2$:
\bea
&&\int d^2\bkappa_1 d^2\bkappa_2 \phi(\bb,x_A,\bkappa_1)
\phi(\bb,x_A,\bkappa_2) {\bkappa_1^2 \over (\bkappa_2^2+m_Q^2)( (\bkappa_1+\bkappa_2)^2+m_Q^2)}
\nonumber\\
&& \approx\pi \int d^2\bkappa_1 \left[\phi(\bb,x_A,\bkappa_1) \phi(\bb,x_A,0) + \phi^{(2)}(\bb,x_A,\bkappa_1)
\right]\log{\bkappa_1^2 \over m_Q^2} \nonumber\\
&& \approx  \pi [\phi(\bb,x_A,0)+\phi^{(2)}(\bb,x_A,0)]
\log{Q_A^2(\bb,x_A) \over m_Q^2}
\label{eq:6.3.6}
\eea
where
$\phi^{(2)}(\bb,x_A,\bkappa)$ is the convolution 
\beq
\phi^{(2)}(\bb,x_A,\bkappa)
=\Big( \phi\otimes \phi \Big)(\bb,x_A,\bkappa)\,.
\label{eq:6.3.7}
\eeq
For heavy nuclei $\phi^{(2)}(\bb,x_A,\bkappa)$ has the same meaning as $\phi(\bb,x_A,\bkappa)$
but for twice larger $T(\bb)$, what counts for the purposes of the present discussion
is that 
\beq
[\phi(\bb,x_A,0)+\phi^{(2)}(\bb,x_A,0)] \approx {1 \over \pi Q_A^2(\bb,x_A)} \propto A^{-1/3}
\label{eq:6.3.8}
\eeq
The final result for the nuclear spectrum at $z\to 1$ is
\bea
&&{(2\pi)^2 d\sigma_A(g^* \to Q\bar{Q}) \over dz d^2\bp d^2\bb}\Biggr|_{z\to 1}
\approx {2T_F\alpha_S(Q_A^2(\bb,x_A)) \over  Q_A^2(\bb,x_A)}\cdot
\log{Q_A^2(\bb,x_A) \over m_Q^2}
\label{eq:6.3.9}
\eea

The first dramatic effect
is that the $\bp$-spectrum of fast quarks will be flat for $\bp^2 \lsim 
 Q_A^2(\bb,x_A)$ which must be contrasted to the $\propto 1/\bp^2$ free-nucleon
spectrum (\ref{eq:6.3.2}). The second dramatic effect is a strong enhancement of
the shadowing effect. The result of the impact parameter integration can be
evaluated as 
\beq
\pi \int d^2\bb[\phi(\bb,x_A,0)+\phi^{(2)}(\bb,x_A,0)] \approx {\sigma_A[\sigma_0(x_A)]
\over \langle Q_A^2(\bb,x_A)\rangle}
\label{eq:6.3.10}
\eeq
where $\langle Q_A^2(\bb,x_A)\rangle$ is an average saturation scale. 
This gives the shadowing effect
\beq
R_{A/N}(z\to 1;x_A,\bp)  \approx {\sigma_A[\sigma_0(x_A)]
\over A\sigma_0(x_A) }\cdot{ \bp^2 \over \langle \langle Q_A^2(\bb,x_A)\rangle \rangle} 
\cdot 
\log{Q_A^2(\bb,x_A) \over m_Q^2}\propto A^{-2/3}\, .
\label{eq:6.3.11}
\eeq
 The analytic formula for the transition from small to large $z$ can be worked
out making use of the explicit parameterization (\ref{eq:6.2.1}). The gross
features of this transition are well reproduced by the interpolation formula
\beq
R_{A/N}(z;x_A,\bp) \approx {\sigma_A[\sigma_0(x_A)] \over A\sigma_0(x_A)}
\left[{\bp^2 \over \bp^2 +z^2\langle Q_A^2(\bb,x_A)\rangle } +
 {z^2  \bp^2 \over \langle Q_A^2(\bb,x_A)\rangle }\cdot 
\log{\langle Q_A^2(\bb,x_A) \rangle \over m_Q^2}\right]
\label{eq:6.3.12}
\eeq
applicable for $\bp^2 \lsim \langle  Q_A^2(\bb,x_A)\rangle $.


\subsection{The Cronin effect: hard quark-jets, $\bp^2 \gsim
Q_A^2(\bb,x_A)$}
\label{Sec:6.4}


\subsubsection{Open charm from the direct and resolved gluon interactions}
\label{Sec:6.4.1}

The above $k_{\perp}$-factorization formulas are exact and can
be directly applied to the calculation of the jet spectra. In
order to make a contact with the more familiar 
treatment of photoproduction of jets, here we present the 
LL$\bp^2$ decomposition of the jet spectra into
the direct and resolved interactions of the incident
parton $a$. Such a decomposition will make more obvious 
the origin of the Cronin effect, although 
in the practical calculations there is no need to
resort to the LL$\bp^2$ approximation.

For hard jets we can use the large-$\bp$ approximation (\ref{eq:B.3}) of
Appendix B. Then the hard single-jet 
spectrum for the free-nucleon target takes the 
form
\bea
&&{d\sigma_N(g^* \to Q\bar{Q}) \over dz d^2\bp}  = \nonumber\\
&&{T_F\alpha_S(\bp^2) [z^2 +(1-z)^2]\over
(2\pi)^2 } \int d^2\bkappa\,\bkappa^2\, f(x_A,\bkappa)\cdot 
\left[{z^2 \over \bp^2} + {(1-z)^2 \over (\bp+\bkappa)^2}\right]\cdot{1 \over
(\bp+z\bkappa)^2}\, , 
\label{eq:6.4.1.1}
\eea
whereas for the heavy nuclear target 
\bea
{d\sigma_A(g^* \to Q\bar{Q}) \over dz d^2\bp d^2\bb}&=&
 {2T_F\alpha_S(\bp^2)[z^2+(1-z)^2]\over (2\pi)^2} \nonumber\\
&\times & \int d^2\bkappa_1 d^2\bkappa_2 \phi(\bb,x_A,\bkappa_1) \phi(\bb,x_A,\bkappa_2)
{(z\bkappa_1 - (1-z)\bkappa_2)^2 \over
(\bp+\bkappa_2)^2 (\bp+z\bkappa_1+z\bkappa_2)^2} \,.\nonumber\\
\label{eq:6.4.1.2}
\eea
For the sake of brevity here we suppressed the contribution from the 
linear term $\propto S_{abs}(\bb)$ in (\ref{eq:5.5}).
 
In the spirit of LL$\bp^2$ one can disentangle two distinct sources of the transverse 
momentum of the quark-jet. In the first case the pQCD subprocess
can be viewed as the large-$\bp$ direct (unresolved)
boson-gluon fusion $g^* g \to Q\bar{Q}$ where
the exchanged gluon $g$ has a small transverse momentum $|\bkappa| \lsim |\bp|$, the
large $\bp$ flows in the $t$-channel of the pQCD subprocess
 $g^* g \to Q\bar{Q}$. Such a contribution can be
evaluated as \bea
&&{d\sigma_N(g^* \to Q\bar{Q}) \over dz d^2\bp}\Biggr|_{dir} =
 {T_F\alpha_S(\bp^2) [z^2+(1-z)^2]^2\over (\bp^2)^2}
 \int^{\bp^2} d^2\bkappa\, \bkappa^2\,f(x_A,\bkappa)\nonumber\\
&& ={T_F\alpha_S(\bp^2) [z^2+(1-z)^2]^2\over (\bp^2)^2} 
\cdot {4\pi^2 \alpha_S(\bp^2)\over N_c}
\cdot
G(x_A,\bp^2)\nonumber\\
&&={2\pi [z^2+(1-z)^2]\over \bp^2}
 \cdot {4\pi^2 \alpha_S(\bp^2)\over N_c}
\cdot {dQ(z,\bp)\over d\bp^2} \cdot G(x_A,\bp^2)\, .
\label{eq:6.4.1.3}
\eea
First, one recovers the familiar collinear 
factorization proportionality to the target gluon density $G(x_A,\bp^2)$.
Second, one readily identifies the unintegrated density of large-$\bp$ quarks in the gluon,
 \beq
{dQ(z,\bp)\over d\bp^2} = {T_F\alpha_S(\bp^2) [z^2+(1-z)^2] \over 2\pi\bp^2}\, .
\label{eq:6.4.1.4}
\eeq
One can say that the direct process probes the transverse momentum
distribution of quarks in the beam gluon $g^*$. 

The second contribution corresponds to the $t$-channel quark being close
to the mass shell, $(\bp+\bkappa)^2 \sim m_Q^2$. In this case the large
transverse momentum $\bp$ of the quark-jet comes from the large transverse
momentum of the exchanged gluon, $\bkappa \approx -\bp$,  in close 
similarity to the pomeron splitting
mechanism for production of hard diffractive dijets \cite{NZsplit}.
It also can be viewed as a jet from the resolved gluon interactions.
All the target information is encoded
in the unintegrated $f(x_A,\bkappa)$ for a free nucleon or $\phi(\bb,x_A,\bp)$
for a nucleus, and the process 
itself looks as the fusion $Qg\to Q'$. One must be careful
with the fusion reinterpretation as the resolved contribution
is only a part pf the more generic $g^*g \to Q\bar{Q}$ and singling it out 
only makes a unique sense to the LL$\bp^2$ approximation.

The case of finite $0 < z < 1$ is a bit more involved and will be 
considered to more detail 
elsewhere. The two poles
in (\ref{eq:6.4.1.1}),
at $(\bp+z\bkappa)^2=0$ and $(\bp+\bkappa)^2=0$, are
well separated. At $z\to 1$ the two poles merge but the residue 
vanishes $\propto (1-z)^2$. Here-below for the sake of 
illustration we focus on slow quark production, $z\ll 1$.
For the massive quarks upon the azimuthal averaging
\bea
\Big\langle {1 \over (\bp-\bkappa)^2+m_Q^2} \Big\rangle \Longrightarrow 
{1\over \sqrt{(\bp^2-\bkappa^2-m_Q^2)^2 +4\bp^2 m_Q^2}}
\Longrightarrow {1 \over |\bp^2-\bkappa^2| +\delta^2}\, ,
\label{eq:6.4.1.5}
\eea
where the infrared regulator of the $d\bkappa^2$ integration
can be taken as
$
\delta^4 \approx \bp^2 m_Q^2\,.
$
The corresponding contribution to the quark-jet cross section
for the free-nucleon target can be evaluated as 
\bea
{d\sigma_N(g^* \to Q\bar{Q}) \over dz d^2\bp}\Biggr|_{res} &\approx &
{T_F\pi \alpha_S(\bp^2)\over
(2\pi)^2} f(x_A,\bp)\log{\bp^2 \over m_Q^2} \nonumber \\
&=&{2\pi\over  \bp^2}
 \cdot {4\pi^2 \alpha_S(\bp^2)\over N_c}
\cdot {d G(x_A,\bp^2)\over d\bp^2}\cdot Q(z\ll 1,\bp)\,.  
\label{eq:6.4.1.7}
\eea
The emerging logarithm times the splitting function can be identified
with the integrated quark structure function $Q(z\ll 1,\bp^2)$ of the beam 
gluon $g^*$ probed by the exchanged gluon $g$ of virtuality $Q_g^2 = \bp^2$. 
The resolved gluon mechanism probes the transverse momentum
distribution of gluons in the target. 

In the more general case
one must take the quark density $Q(x_B,\bp^2)$ where $x_B$ is the 
fraction of the lightcone momentum of the beam carried by the quark
$Q$.  Combining together (\ref{eq:6.4.1.3}) and
(\ref{eq:6.4.1.7}) we find the spectrum of hard quark-jets which is
proportional to 
\beq
\left({ dQ(x_B,\bp^2)\over d\bp^2} G(x_A,\bp^2) + 
Q(x_B,\bp^2){d G(x_A,\bp^2) \over d\bp^2}\right)=
{ d \over d\bp^2}[Q(x_B,\bp^2) G(x_A,\bp^2)]\, ,
\label{eq:6.4.1.8}
\eeq
which restores the beam-target symmetry. 
A similar observation has been made in \cite{Kharzeev:2003wz}. The last form shows that in 
the general case the both mechanisms are of comparable importance. 
It also demonstrates the nonlocality
properties of the $\bar{b}\Pom \bar{b}\Pom$ vertex in the Kancheli-Mueller
diagram of Fig.~\ref{fig:SingleJetKancheli_N}d.
The derivation of the nuclear counterparts of (\ref{eq:6.4.1.3}) 
and  (\ref{eq:6.4.1.7}) 
for $z\ll 1$ is straightforward and need not be repeated here. 
We proceed directly to the derivation of the Cronin effect.


\subsubsection{Antishadowing property of collective nuclear gluon density
as an origin of the Cronin effect}
\label{Sec:6.4.2}

For a more accurate  isolation of the Cronin effect for hard jets consider the
nuclear excess cross section
\beq
{d\Delta \sigma_A \over d^2\bb} = {d\sigma_A \over d^2\bb} - {1\over 2}T(\bb)d\sigma_N \, .
\label{eq:6.4.2.1}
\eeq
Again we start with $z\ll 1$. Suppressing common 
factors and making use of (\ref{eq:6.4.1.5}) we need to evaluate
\bea
{d\Delta\sigma_A \over dz d^2\bp d\bb} \propto
\int d^2\bkappa{\bkappa^2 \over |\bp^2-\bkappa^2| +\delta^2}[\phi(\bb,x_A,\bkappa)-
{1\over 2}T(\bb)f(x_A,\bkappa)]
\label{eq:6.4.2.2}
\eea
A priori the sign of this quantity is not obvious because the 
function $\Delta\phi_A(\bb,x_A,\bkappa) = \phi(\bb,x_A,\bkappa)-
{1\over 2}T(\bb)f(x_A,\bkappa)$ is negative valued for $\bkappa^2 \lsim Q_A^2(x_A,\bb)$
and positive valued in the antishadowing region described by eq.~(\ref{eq:6.4.1.2}).
We decompose $\bkappa^2 /( |\bp^2-\bkappa^2| +\delta^2)$ as follows:
\bea
 {\bkappa^2 \over  |\bp^2-\bkappa^2| +\delta^2} 
  = \theta(\bp^2-\bkappa^2)\cdot{\bkappa^2 \over \bp^2} 
  + \theta(\bkappa^2-\bp^2) + K(\bp^2,\bkappa^2)\,.
\label{eq:6.4.2.3}
\eea
Here the first term isolates the direct boson-fusion contribution, whereas
the collinear quark-pole contribution is contained entirely in 
$K(\bp^2,\bkappa^2)$ which can readily be shown to be a positive-valued 
function which vanishes $\sim \kappa^4$ for $\bkappa^2 \ll \bp^2$ and
$\sim 1/\bkappa^2$ for $\bkappa^2 \gg \bp^2$. 

The shadowing effect from the negative-valued $\Delta\phi_A(\bb,x_A,\bkappa)$ at
small $\bp^2 \lsim Q_A^2(\bb,x_A)$ is concentrated in the  contribution from 
the first term in (\ref{eq:6.4.2.3}) which by virtue of the sum rule 
(\ref{eq:6.2.6}) can be represented as  
\beq
{1\over \bp^2}\int_0^{\bp^2} d\bkappa^2\, \bkappa^2\, \Delta\phi_A(\bb,x_A,\bkappa) =
-{1\over \bp^2}\int_{\bp^2}^{\infty} d\bkappa^2\, \bkappa^2\, \Delta\phi_A(\bb,x_A,\bkappa)\, .
\label{eq:6.4.2.4}
\eeq
Consequently, the nuclear excess cross section is entirely calculable in
terms of gluon densities in the hard region of $\bkappa^2 \gsim \bp^2$: 
\bea
\bp^2{d\Delta\sigma_A \over dz d^2\bp d\bb} \propto
&-&{1\over \bp^2}\int_{\bp^2}^{\infty} d\bkappa^2\, \bkappa^2\, \Delta\phi_A(\bb,x_A,\bkappa)
\nonumber\\
&+&\int_{\bp^2}^{\infty} d\bkappa^2\,  \Delta\phi_A(\bb,x_A,\bkappa)\nonumber\\
&+&{1\over \pi}\int d^2\bkappa\, K(\bp^2,\bkappa^2) \Delta\phi_A(\bb,x_A,\bkappa)
\label{eq:6.4.2.5}
\eea
 Now notice that in this hard region 
$
\Delta\phi_A(\bb,x_A,\bkappa) \propto  1/(\bkappa^2)^3 
$, see eq.~(\ref{eq:6.2.3}), 
and all integrals in (\ref{eq:6.4.2.5}) are well converging ones. The collinear
quark-pole contribution can be evaluated as it was done in (\ref{eq:6.4.1.7}),
factoring  out $\bp^2\Delta\phi_A(\bb,\bp)$, so that to logarithmic
accuracy  
\bea
\bp^2{d\Delta\sigma_A \over dz d^2\bp d\bb} &\propto&
\Delta\phi_A(\bb,\bp)\left[-1+{1\over 2} + \log{\bp^2 \over m_Q^2}\right] \,.
\label{eq:6.4.2.6}
\eea
Following the derivation of eq.~(\ref{eq:6.4.1.7}) we can associate
$\log(\bp^2/m_Q^2)$ with $ Q(z\ll 1,\bp)$ and to LL$\bp^2$ accuracy
\bea
{d\Delta\sigma_A \over dz d^2\bp d\bb} & =& {d\sigma_N \over dz d^2\bp}\Biggr|_{res}
\cdot T(\bb)\cdot\left[{f_A(\bb,x_A,\bp^2) \over f(x_A,\bp^2)} -1\right]\nonumber\\
&=&
{d\sigma_N \over dz d^2\bp}\Biggr|_{res}\cdot T(\bb) \Delta_{HT}(\bb,x_A,\bp).
\label{eq:6.4.2.7}
\eea

The result for the resolved interactions does not depend on the projectile
parton $a$, 
\bea
[R_{A/N}(z\ll 1;x_A,\bp)-1]_{res} =  
\Delta_{HT}(\bb,x_A,\bp)\,.
\label{eq:6.4.2.8}
\eea
while the dilution of the Cronin effect by the contribution from direct interactions to 
the free-nucleon is process dependent:
\bea
R_{A/N}(z\ll 1;x_A,\bp)-1 =
{d\sigma_N(res) \over d\sigma_N(res)+ d\sigma_N(dir)}\cdot [R_{A/N}(z\ll 1;x_A,\bp)-1]_{res} \, .
\label{eq:6.4.2.9}
\eea
Hereafter we focus on the projectile independent antishadowing for resolved interaction
and suppress the subscript '$res$'. For heavy nuclei 
\beq
\int d^2\bb \, T^2(\bb) \approx {9A^2 \over 8\pi R_A^2}
\label{eq:6.4.2.10}
\eeq
and we obtain our final estimate for the antishadowing Cronin effect
at large-$\bp^2$ 
\beq
\Delta_{Qg}(z\ll 1;x_A,\bp)=R_{A/N}(z\ll 1;x_A,\bp)-1 \approx  {9\pi\gamma^2 \over 4N_c} \cdot 
{\alpha_S(\bp^2)\cdot G(x_A,\bp^2) \over \bp^2} \cdot {A\over R_A^2}\,.
\label{eq:6.4.2.11}
\eeq
Here we adopted the same subscript $Qg$ as in the familiar splitting
functions. 

First, of the two components the direct $g^*g$--fusion would 
have given nuclear shadowing, see the first line in 
eq.~(\ref{eq:6.4.2.5}), the antishadowing nuclear
 excess cross section is a feature 
of the resolved gluon interactions.
Second, the fundamental point is that for hard jets 
the nuclear excess cross section 
follows directly from the nuclear antishadowing component 
$\Delta_{HT}(\bb,x_A,\bp)$ of the collective nuclear gluon density
\cite{NSSdijet,NSSdijetJETPLett}. 
Third,  the antishadowing rises with the nuclear mass number,
$R_{A/N}(z;x_A,\bp)-1 \propto A^{1/3}$. 
Fourth, the antishadowing nuclear excess is a special quadratic 
functional of the gluon density in the proton -
a product of the integrated and unintegrated gluon densities.
Finally, for hard jets the antishadowing Cronin effect vanishes 
$\propto 1/p^2$. Whereas the antishadowing
for hard jets, $R_{A/N}(z;x_A,\bp) > 1$, has been discussed by many authors
under model assumptions and specific parameterizations for the
dipole cross section/unintegrated glue (e.g. 
\cite{BaierKovWied,JalilianRaju,Kharzeev:2003wz,Iancu:2004bx}
and references therein),
the above reported model independent derivation, and 
identification of the source,
of antishadowing 
Cronin effect are new results. 
 

\subsubsection{Variations of the Cronin effect from slow to leading
charm}
\label{Sec:6.4.3}

The convolution representation (\ref{eq:6.1.4}) makes it obvious that
the above findings on the Cronin effect are directly applicable to
the opposite limiting case of $z\to 1$ as well. We recall that
the free-nucleon spectra for $z\to 1$ and $z\ll 1$ are identical. 
The smearing of the
decreasing spectrum (\ref{eq:6.1.3}) does obviously enhance the 
antishadowing effect at large $\bp$ which can readily be seen as
follows. The $\bp$-dependence of the spectra (\ref{eq:6.4.1.3})
and (\ref{eq:6.4.1.7}) is driven by the factor $1/( \bp^2)^2$ and the
enhancement of the Cronin effect will for the most part be due
to the smearing of $1/[ (\bp-\bkappa)^2]^2$ in (\ref{eq:6.1.4}). 
The large-$\bp$ expansion
\beq
{ 1\over [ (\bp-\bkappa)^2]^2} \Rightarrow {1\over ( \bp^2)^2}
\left[1+ 4 {\bkappa^2 \over \bp^2}\right]\, ,
\label{eq:6.4.3.1}
\eeq
in which the azimuthal averaging is understood, gives an
enhancement of $R_{A/N}(z\ll 1;x_A,\bp)$ by the extra factor
\bea
\rho_{A/N}(\bp) &=& \int d^2\bkappa \phi(\bb,x_A,\bkappa)\left[1+ 4 {\bkappa^2 \over \bp^2}\right]
\nonumber\\
&=& 1+ 2 
{\alpha_S(\bp^2)G(x_A, \bp^2) \over \alpha_S(Q_A^2)G(x_A,Q_A^2)}
\cdot {Q_A^2(\bb,x_A) \over \bp^2} \approx 1+\Delta_{HT}(\bb,x_A,\bp)\,,
\label{eq:6.4.3.2}
\eea
cf. eq.~(\ref{eq:6.2.3}) for $\gamma\approx 2$. Consequently, at $z \to 1$ the
antishadowing Cronin effect will be twice stronger than that at $z\ll 1$,
\beq
\Delta_{Qg}(z\to 1;x_A,\bp) =R_{A/N}(z\ll 1;x_A,\bp)\cdot \rho_{A/N}(\bp) -1
\approx 2 \Delta_{Qg}(z\ll 1;x_A,\bp)\, ,
\label{eq:6.4.3.3}
\eeq
what nicely correlates with the
stronger nuclear shadowing at small $\bp$. The same correlation is obvious
from the sum rule which relates the $\bp$-integrated nuclear cross sections
for $z\to 1$ and $z\ll 1$,
\bea
{ d\sigma_A(g^* \to Q\bar{Q}) \over dz d^2\bb}\Biggr|_{z\to 1 }=
{ d\sigma_A(g^* \to Q\bar{Q}) \over dz d^2\bb}\Biggr|_{z\ll 1 }\,,
\label{eq:6.4.3.4}
\eea
which readily follows from the convolution representation (\ref{eq:6.1.4}).


\subsubsection{Numerical estimates for the Cronin effect}
\label{6.4.4}

\begin{figure}[!t]
\begin{center}
\includegraphics[width = 13.0cm]{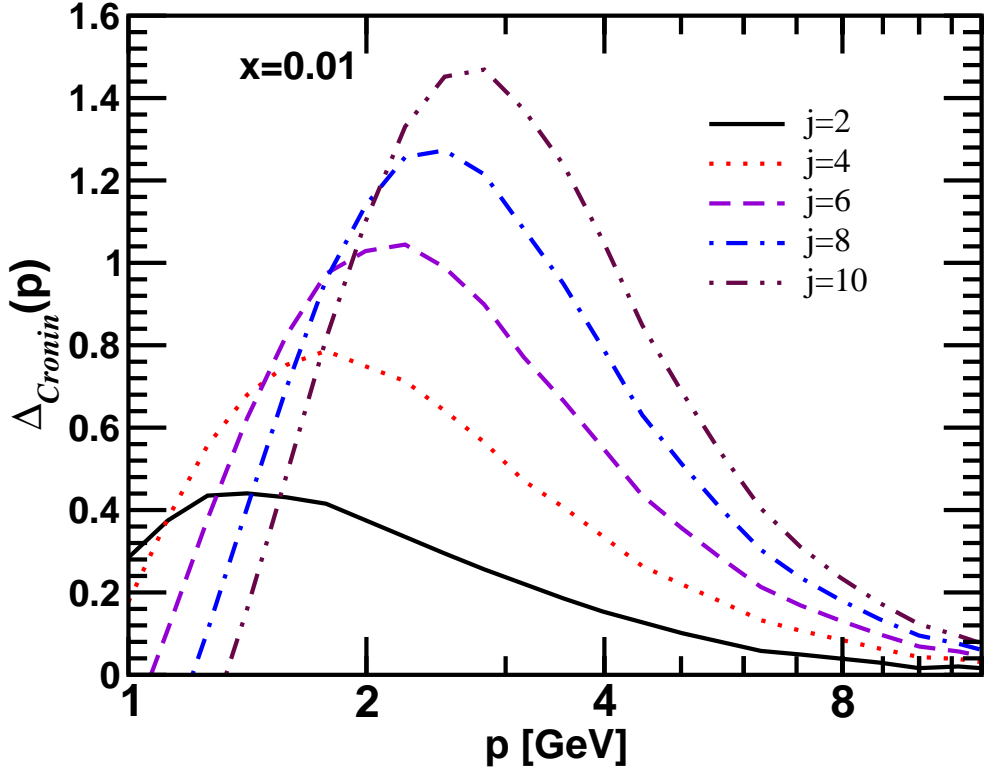}
\caption{
(Color online)
The antishadowing Cronin effect in jet production 
off a heavy nucleus at $x = 0.01$ as a function of the transverse
momentum $p$ of the jet parton for several multiplicities of
overlapping nucleons $j$ in the Lorentz-contracted nucleus. 
The production of slow quarks off
nucleus with mass number $A=200$ corresponds to $j=4$; the
production of slow gluons off nucleus with mass number $A=200$
corresponds to $j=8$. }
\label{fig:DeltaCronin1}
\end{center}
\end{figure}

\begin{figure}[!t]
\begin{center}
\includegraphics[width = 13.0cm]{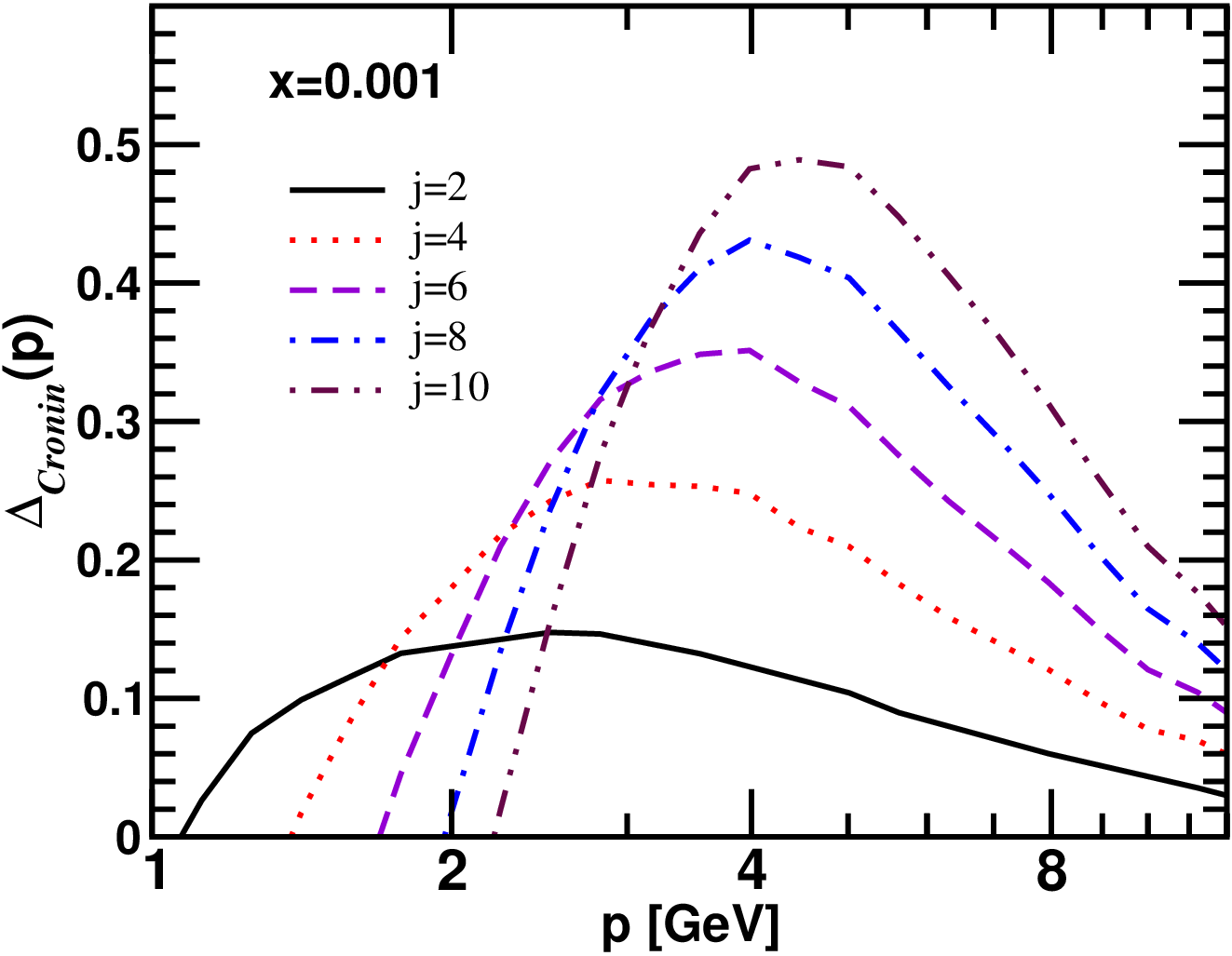}
\caption{
(Color online)
The antishadowing Cronin effect in jet production 
off a heavy nucleus at $x=0.001$ as a function of the transverse
momentum $p$ of the jet parton for several multiplicities of
overlapping nucleons $j$ in the Lorentz-contracted nucleus. 
The production of slow quarks off
nucleus with mass number $A=200$ corresponds to $j=4$; the
production of slow gluons off nucleus with mass number $A=200$
corresponds to $j=8$. }
\label{fig:DeltaCronin2}
\end{center}
\end{figure}

The antishadowing properties of the collective nuclear gluon
density have been studied  \cite{NSSdijet,NSSdijetJETPLett} in
connection with the $A$-dependence of coherent diffractive
dijet production.
The starting point was 
the unintegrated gluon density
in the proton, $f(x,\bkappa)$, adjusted to reproduce the $\gamma^*N$
interactions from real photoproduction, $Q^2=0$, to DIS at large
$Q^2$ \cite{INDiffGlue}.  The quantity of interest is 
the antishadowing higher twist effect in the collective nuclear
gluon density for $j$ overlapping nucleons which is given by the
$j$-fold convolution  $f^{(j)}(x_A,\bkappa)$. This density per
overlapping nucleon was represented in ref. 
\cite{NSSdijet}  as
\bea
f_A(\bb,x_A,\bkappa) &=& {f^{(j)}(x_A,\bkappa) \over j \sigma_0^{j-1}}
= 
f(x_A,\bkappa )\left[1+ (j-1)\Delta_j(x_A,\bkappa)\right]\,.
\label{eq:6.4.4.1}
\eea
With reference to the Poissonian form of the multiple scattering expansion 
(\ref{eq:6.2.5}), 
in Figs.~\ref{fig:DeltaCronin1},\ref{fig:DeltaCronin2} we show
$\Delta_{Cronin}(x_A,\bkappa)=j \cdot \Delta_j(x_A,\bkappa)$ for two
values of $x_A$. 

These results for $\Delta_{Cronin}(x_A,\bkappa)$
can be used to evaluate 
$\Delta(x_A,\bp)= R_{A/N}(x_A,\bp)-1$ for different jet
production subprocesses and for different nuclei:
\bea
\Delta_{Qg}(z\ll 1;x_A,\bp) \approx \Delta_{Cronin}(x_A,\bkappa)
\Big|_{j\approx \langle j \rangle}
\label{eq:6.4.4.2}
\eea
For the $^{196}Pt$ nucleus, which is not
any different from gold - the favorite of RHIC, 
the average value of $j$ in
the expansion (\ref{eq:6.2.5}) is $\langle j\rangle 
\approx 4$. 
At $x_A=0.01$ the peak value of the antishadowing effect,
$\Delta_{Qg}(z\ll 1;x_A,\bp) \approx 0.7$,  
will be reached at $p\approx (2\div 2.5)$ GeV. 
It is important that although the numerical studies in
\cite{Nonlinear} suggest not so large $Q_A^2 \lsim 1$ GeV$^2$,
the peak of the antishadowing effect is placed 
at transverse momenta
$p$ which are sufficiently large to justify the pQCD evaluations.
For quark jets with $z\to 1$ the antishadowing effect will be twice stronger. 
The relationship between $\Delta_{Cronin}(x_A,\bkappa)$ and 
$\Delta_{gg}(z;x_A,\bp), \Delta_{gq}(z;x_A,\bp)$ 
for gluon jets will be discussed in Sections \ref{Sec:7.2} and \ref{Sec:8.2}, 
respectively.

The $x_A$-dependence of the Cronin effect is noteworthy.
At $x_A=0.001$ the peak value of the antishadowing effect,
$\Delta_{Qg}(z\ll 1; x_A,\bp) \approx 0.25$, 
will be reached at $p\approx (3\div 3.5)$ GeV.
This $x$-dependence of the position and magnitude of the peak value of
the antishadowing effect is driven by three effects.
The first one is the $x$-dependence 
of $Q_A^2(x,\bb)$ which is similar to that of
the real photoabsorption cross section and is rather weak, $\sim x^{-0.1}$.
The second source is the variation of the 
$\bkappa^2$-dependence of  $f(x,\bkappa)$ as $x$ decreases.
Indeed, the phenomenological studies of ref. \cite{INDiffGlue}
show that while the exponent $\gamma\approx 2$ is appropriate for $x_A=0.01$, at
$x_A=0.001$ the $\bkappa^2$-dependence of the unintegrated gluon density
in the region of $\bkappa^2$ of interest 
corresponds to $\gamma \approx 1.7$ which lowers substantially the
magnitude, and shifts to larger $\bkappa^2$ the onset of the antishadowing 
regime (\ref{eq:6.2.3}). Finally, because of the shift to larger $\bp^2$ 
with decreasing $x$ the peak value of $\Delta_{Cronin}(\bp)$
is suppressed by the $1/\bp^2$ dependence of the 
antishadowing contribution, see eq.~(\ref{eq:6.4.2.11}). This explains the numerical
findings of Ref.~\cite{NSSdijet}.

We recall that the shown numerical results are only for the resolved 
interactions, any comparison with the experimental data requires an
application of the dilution factor (\ref{eq:6.4.2.9}). Also, the
momentum of the observed hadrons is only a fraction of the jet
momentum what places the Cronin peak for hadrons at smaller values
of $\bp$ than for jets. 


\section{Nonlinear $k_{\perp}$-factorization for production of gluons
off nuclei}
\label{Sec:7}


\subsection{The gluon spectra for the free-nucleon and nuclear targets}
\label{Sec:7.1}

The excitation of the gluon $g \to gg$ offers a still more remarkable
example of breaking of linear $k_{\perp}$-factorization. In this case
$a=b=c=g$ and  we take the shorthand notations
$z\equiv z_g$ and $\bp\equiv \bp_g$. The 3-parton cross section for generic three-gluon
state equals
\bea
\sigma_{ggg}(\brho,\br)={C_A \over 2C_F}[\sigma_{q\bar{q}}(x,\brho)+\sigma_{q\bar{q}}(x,\brho-\br)
+\sigma_{q\bar{q}}(x,\br)]
\label{eq:7.1.1}
\eea
The dipole cross section structure 
in eq.~(\ref{eq:3.2}) takes the form
\bea
{C_A\over
2C_F}\Big\{ &&\sigma_{q\bar{q}}(x,\br) + \sigma_{q\bar{q}}(x,z\br')+ \sigma_{q\bar{q}}(x,\br-z\br')\nonumber\\
+&&\sigma_{q\bar{q}}(x,\br') + \sigma_{q\bar{q}}(x,z\br)+ \sigma_{q\bar{q}}(x,\br'-z\br) \nonumber\\
-&&2\sigma_{q\bar{q}}(x,\br-\br') -2 \sigma_{q\bar{q}}(x,z(\br-\br')) \Big\} 
\label{eq:7.1.2}
\eea 
Repeating the analysis of sections \ref{Sec:3} and \ref{Sec:4}, for the
free-nucleon case one readily finds the $k_{\perp}$-factorization
representation for the gluon-jet cross section
\bea 
{d\sigma_N(g^*
\to gg) \over dz d^2\bp} &=& {1 \over 2
(2\pi)^2} \int d^2\bkappa f(x_A,\bkappa) \times  {C_A\over
2C_F} \Big\{|\Psi(z,\bp) -\Psi(z,\bp+z\bkappa)|^2 \nonumber\\
&+& |\Psi(z,\bp+\bkappa)
-\Psi(z,\bp+z\bkappa)|^2\
 +|\Psi(z,\bp) -\Psi(z,\bp+\bkappa)|^2 \Big\}\, .
\label{eq:7.1.3} 
\eea
In agreement with the generic arguments of section \ref{Sec:4},
the abelian limit is recovered for $z\ll 1$: \bea {d\sigma_N(g^*
\to gg) \over dz d^2\bp} \Biggr|_{z\ll 1}= {1 \over 2
(2\pi)^2} \int d^2\bkappa f(x_A,\bkappa) \,
{C_A\over C_F} \cdot|\Psi(z,\bp) -\Psi(z,\bp+\bkappa)|^2 \, .
\label{eq:7.1.4}
\eea
The same abelian limit is recovered also for $z\to 1$, but this
equality of two limiting cases will again be upset for nuclear targets.

Following the discussion in section \ref{Sec:6.4}, one can readily 
decompose the spectrum (\ref{eq:7.1.4}) into the contributions
from the direct and resolved gluon interactions, we shall skip
those details here.

The excitation of gluons into gluons, $g\to gg$, is an exceptional case
for which exact formulas for a nuclear target derive without invoking
the large-$N_c$ approximation.
Our result for the nuclear gluon jet spectrum reads
\bea &&{(2\pi)^2 d\sigma_A(g^* \to gg) \over dz d^2\bp d^2\bb}=  
\left[S_{abs}^{(g)}(\bb) \right]^2\cdot \int d^2\bkappa \phi_g(\bb,x_A,\bkappa) \nonumber\\
&&\times \Big\{|\Psi(z,\bp) -\Psi(z,\bp+z\bkappa) |^2 +
|\Psi(z,\bp +\bkappa) -\Psi(z,\bp+z\bkappa)|^2 + |\Psi(z,\bp) -\Psi(z,\bp+\bkappa) |^2 \Big\}
\nonumber \\
&&+  S_{abs}^{(g)}(\bb) \int d^2\bkappa_1 d^2\bkappa_2 \phi_g(\bb,x_A,\bkappa_1)
\phi_g(\bb,x_A,\bkappa_2)\nonumber\\
&&\times \Big\{|\Psi(z,\bp+\bkappa_1) -\Psi(z,\bp+z\bkappa_2) |^2 +
|\Psi(z,\bp +\bkappa_1+\bkappa_2) -\Psi(z,\bp+z\bkappa_2)|^2 \nonumber
\\
&& + |\Psi(z,\bp+\bkappa_1) -\Psi(z,\bp+z(\bkappa_1+\bkappa_2)) |^2 \Big\}\nonumber\\
&&+\int d^2\bkappa_1 d^2\bkappa_2 d^2\bkappa_3
\phi_g(\bb,x_A,\bkappa_1) \phi_g(\bb,x_A,\bkappa_2) \phi_g(\bb,x_A,\bkappa_3)\nonumber\\
&&\times
|\Psi(z,\bp+\bkappa_1+\bkappa_3)
-\Psi(z,\bp+z(\bkappa_2+\bkappa_3))|^2 \, . 
\label{eq:7.1.5} \eea
Here $\phi_g(\bb,x_A,\bkappa)$ is 
the collective nuclear glue  which now must be evaluated
from
\bea \Gamma_{A}[\bb,{C_A \over 2C_F}\sigma_{q\bar{q}}(x_A,\br)] \equiv \int
d^2\bkappa \phi_g(\bb,x_A,\bkappa) \Big[1 - \exp[i\bkappa\br] \Big] \, ,\nonumber\\
S_{abs}^{(g)}(\bb)= \exp[-{1\over 2}\cdot {C_A\over 2C_F}\cdot\sigma_0(x_A) T(\bb)]\,.
\label{eq:7.1.6}
\eea
in which the nonabelian factor $C_A/2C_F$ enters manifestly.
The emergence of a new collective gluon density 
$\phi_g(\bb,x_A,\bkappa) \neq \phi(\bb,x_A,\bkappa)$ illustrates nicely
the point made in \cite{Nonlinear} that the collective nuclear
glue must be described by a density matrix in color space
rather than by a universal scalar function. Correspondingly, the
nuclear pomeron $\Pom_A^g$ associated with $\phi_g(\bb,x_A,\bkappa)$
is different from $\Pom_A$ defined for charm production off nuclei
in Sect. \ref{Sec:5}.
The saturation scale
for the gluon density $\phi_g(\bb,x_A,\bkappa)$ differs from
$Q_A^2(\bb,x_A)$  by the factor $C_A/2C_F$.
Also, for hard gluons
\bea
f_A^{(g)}(\bb,x_A,\bkappa) = 
 f(x_A,\bkappa )\cdot {C_A \over 2C_F}\cdot 
\left[1+ {C_A \over 2C_F}\Delta_{HT}(\bb,x_A,\bkappa)\right] \,,
\label{eq:7.1.7}
\eea

In close similarity to open charm production, the first term in eq.(\ref{eq:7.1.5}) 
which satisfies
linear $k_{\perp}$-factorization (\ref{eq:7.1.3}) subject to the substitution
of the free-nucleon $f(x_A,\bkappa)$ by the collective nuclear glue
$\phi_g(\bb,x_A,\bkappa)$, is suppressed by the square of the nuclear absorption
factor, $\left[S_{abs}^{(g)}(\bb)\right]^2$. The second term, which is quadratic in the nuclear
gluon density $\phi_g(\bb,x_A,\bkappa)$, is suppressed by $S_{abs}^{(g)}(\bb)$.
What survives for strongly absorbing nuclei is the third term which is
a cubic functional of collective nuclear glue $\phi_g(\bb,x_A,\bkappa)$
and the pattern of
$k_{\perp}$-factorization for single gluon jet production changes
from the conventional linear one for peripheral impact parameters
to the cubic one for small impact parameters. The three components of 
the gluon spectrum can be associated with the three Kancheli-Muller
diagrams of Fig.~\ref{fig:SingleJetKancheli_A}. 
A comparison of (\ref{eq:7.1.3}) and (\ref{eq:7.1.5}) shows that
the nuclear cross section cannot be represented as an expansion in
multiple convolutions of the free nucleon cross section 
as was suggested in \cite{Accardi}.
The $z$-dependence
of the integrands of the quadratic and cubic terms is manifestly 
different from that of the free-nucleon cross section. The
difference between the the free-nucleon and nuclear gluon densities 
entails different $z$-distributions for free-nucleon and nuclear
targets even in the linear term. Consequently, all the three
components of the nuclear spectrum will exhibit the 
$\bp$-dependent LPM effect.

The cubic nonlinearity in terms of the collective gluon density is a
pertinent feature of the radiation of gluons with finite $z$.
Some interesting simplifications are found in the limiting case of $z\ll 1$.
In this case in the quadratic and cubic terms one will encounter the quantities of the
from
\bea
\int d^2\bkappa_1 d^2\bkappa_2 \phi_g(\bb,x_A,\bkappa_1)
\phi_g(\bb,x_A,\bkappa_2)
|\Psi(z,\bp +\bkappa_1+\bkappa_2) -\Psi(z,\bp)|^2 \nonumber\\
=\int d^2\bkappa  \phi_g^{(2)}(\bb,x_A,\bkappa) |\Psi(z,\bp +\bkappa)
-\Psi(z,\bp)|^2 \,. 
\label{eq:7.1.8}
\eea
In terms of the convolution $\phi_g^{(2)}(\bb,x_A,\bkappa)$ 
the resulting single jet spectrum can be cast in the
form
\bea &&{(2\pi)^2 d\sigma_A(g^* \to g g) \over dz d^2\bp
d^2\bb}\Biggr|_{z\ll 1}= \nonumber\\
&&\int d^2\bkappa
\left[2S_{abs}^{(g)}(\bb)\phi_g(\bb,x_A,\bkappa) +
\phi_g^{(2)}(\bb,x_A,\bkappa)\right] \cdot |\Psi(z,\bp +\bkappa)
-\Psi(z,\bp)|^2\, ,
\label{eq:7.1.9} \eea
and the triple-$\Pom_A^g$ exchange Kancheli-Mueller diagram 
of Fig.~\ref{fig:SingleJetKancheli_A}c simplifies to the
double-$\Pom_A^g$ exchange diagram 
of Fig.~\ref{fig:SingleJetKancheli_A}b. 
Finally, we observe that
\beq \phi_{gg}(\bb,x_A,\bkappa) =
2S_{abs}^{(g)}(\bb)\phi_g(\bb,x_A,\bkappa) + \phi_g^{(2)}(\bb,x_A,\bkappa)
\label{eq:7.1.10} 
\eeq 
is precisely still another collective nuclear gluon
density defined in terms of the gluon-gluon dipole cross section:
\beq
 \Gamma_{A}[\bb,{C_A \over C_F}\sigma_{q\bar{q}}(x,\br)] \equiv \int
d^2\bkappa \phi_{gg}(\bb,x,\bkappa) \Big[1 - \exp[i\bkappa\br] \Big]
\label{eq:7.1.11} 
\eeq
To this end,
eq.~(\ref{eq:7.1.9}) is analogous to the linear
$k_{\perp}$-factorization for the free nucleon case, see
(\ref{eq:7.1.4}), but the free nucleon glue $f(x_A,\bkappa)$ is
substituted by $\phi_{gg}(\bb,x_A,\bkappa)$, which is still
a nonlinear --quadratic-- functional of the collective nuclear glue
$\phi_{g}(\bb,x_A,\bkappa)$, and with which one must associate 
still another nuclear pomeron $\Pom_A^{gg}$. 
In terms of  $ \phi_{gg}(\bb,x_A,\bkappa)$
the transverse momentum spectrum of gluon jets with $z\ll 1$
satisfies the linear $k_{\perp}$--factorization and is 
described by the Kancheli-Mueller
diagram of Fig.~\ref{fig:SingleJetKancheli_A}a with $\Pom \to \Pom_A^{gg}$. 
Our final result for $z\ll 1$ coincides 
in its color dipole form with the
one cited in \cite{KovchegovMueller},
see also a discussion in \cite{Nonlinear}, all the results for finite $z$ are new.


\subsection{The Cronin effect for gluons}
\label{Sec:7.2}

The implications of (\ref{eq:7.1.11}) for the Cronin effect 
for gluon-jets are straightforward. The principal change is 
that for soft gluons the collective nuclear
gluon density $\phi_{gg}(\bb,x_A,\bkappa)$ is obtained from 
$\phi(\bb,x_A,\bkappa)$ by the substitution
\beq
Q_{A}^2(\bb,x_A) \Longrightarrow Q_{A,gg}^2(\bb,x_A) \approx {C_A\over C_F}Q_{A}^2(\bb,x_A)\,,
\label{eq:7.2.1} 
\eeq
so that the plateau in $\phi_{gg}(\bb,x_A,\bkappa)$ will be broader than 
in $\phi(\bb,x_A,\bkappa)$ by the factor $C_A/ C_F$. For hard gluons
\bea
f_A^{(gg)}(\bb,x_A,\bkappa) = 
 f(x_A,\bkappa )\cdot {C_A \over C_F}\cdot 
\left[1+ {C_A \over C_F}\Delta_{HT}(\bb,x_A,\bkappa^2)\right] \,,
\label{eq:7.2.2}
\eea
Nuclear shadowing for $\bp^2 \lsim  Q_{A,gg}^2(\bb,x_A)$ will be 
stronger than for the quark jets:
\beq
R_{A/N}^{g}(z;x_A,\bp) = {\sigma_A[{C_A\over C_F}\sigma_0(x_A)] \over A\sigma_{0}(x_A)}
\sim  {C_F \over C_A}R_{A/N}^{Q}(z;x_A,\bp)
\label{eq:7.2.3} 
\eeq
Evidently the antishadowing effect $\Delta_{gg}(z;x_A,\bp)>0 $ at $\bp^2 \gsim  
Q_{A,gg}^2(\bb,x_A)$ will persist for gluon-jets too. Following the 
analysis of section \ref{Sec:6.4}, one would readily find that the nuclear
antishadowing effect will be dominated by the resolved gluon 
interactions, and 
\beq
\Delta_{gg}(z;x_A,\bp)\approx  {C_A\over C_F} \Delta_{Qg}(z;x_A,\bp) \, .
\label{eq:7.2.4}
\eeq
In view of the
larger saturation scale (\ref{eq:7.2.1}) the position of the peak value
of the antishadowing effect for gluon jets, $\Delta_{gg}(z;x_A,\bp)$, 
will be placed at higher $\bp^2$ 
than that for quark jets. For semiquantitative estimates we
again invoke the numerical results for $\Delta_{Cronin}(x_A,\bp)$ shown in Fig.~4
of Ref. \cite{NSSdijet}. For heavy nuclei like gold
or platinum  
$\langle j\rangle_{g} \approx {C_A\over C_F}\langle j \rangle \approx 9$.
At $x_A = 0.01$ quite a large peak value $\Delta_{gg}(z;x_A,\bp) \approx 1.3$ will be
reached at $p \approx 3$ GeV.
At $x_A = 0.001$ the peak value $\Delta_{gg}(z;x_A,\bp) \approx 0.45$ will be
reached at $p \approx 5$ GeV. Evidently, the fragmentation of jets will shift 
the peak position for observed hadrons to lower values of $\bp$. 
The antishadowing properties
of the spectrum of leading gluons with $z\to 1$ are readily derived
following the discussion of quark jets in
section \ref{Sec:6.4}. The full treatment 
of virtual corrections to the production of leading gluons
from the lower order pQCD processes
is beyond the scope of the present communication,
here we only notice that they are arguably small 
if the transverse momentum $\bp$ of the observed leading gluon  
is much larger than the transverse momentum of the beam-gluon $g^*$. 
The leading gluon version of eq.~(\ref{eq:7.1.5}) demonstrates 
how the
pertinent nonlinearity at $z\to 1$ changes the antishadowing
Cronin effect compared to the above discussed case of $z\ll 1$. 
Define first the two special nuclear spectra
\bea 
&&{(2\pi)^2 d\sigma_A^{(g1)} \over dz d^2\bp d^2\bb} =
 \int d^2\bkappa \phi_g(\bb,x_A,\bkappa)|\Psi(z,\bp) -\Psi(z,\bp+\bkappa) |^2 \, \\
\label{eq:7.13}
&&{(2\pi)^2 d\sigma_A^{(g2)}\over dz d^2\bp d^2\bb} = 
\int d^2\bkappa_1 d^2\bkappa_2 
\phi_g(\bb,x_A,\bkappa_1) \phi_g(\bb,x_A,\bkappa_2) |\Psi(z,\bp+\bkappa_1)
-\Psi(z,\bp+\bkappa_2)|^2 \, , \nonumber\\
\label{eq:7.2.5}
\eea
in terms of which
\bea
 &&{ d\sigma_A(g^* \to gg) \over dz d^2\bp d^2\bb} 
\Biggr|_{z\to 1}= \nonumber\\
&& 2S_{abs}^2(\bb) 
{ d\sigma_A^{(1)}(g^* \to gg) \over dz d^2\bp d^2\bb} +
2 S_{abs}(\bb) { \big(d\sigma_A^{(g1)}\otimes \phi_g\Big)(\bp)
\over dz d^2\bp d^2\bb}
+ { \big(d\sigma_A^{(g2)}\otimes \phi_g\Big)(\bp)
\over dz d^2\bp d^2\bb}\, .
\label{eq:7.2.6}
\eea
The small and large-$\bp$ properties of the first two terms in
(\ref{eq:7.2.6}) have already been studied in section \ref{Sec:6}. 
After averaging over the azimuthal angles of $\bkappa_{1,2}$:
\bea
&&\Big\langle \Big\langle |\Psi(z,\bp+\bkappa_1)
-\Psi(z,\bp+\bkappa_2)|^2 \Big\rangle \Big\rangle = \nonumber\\
&=& 
\Big\langle {1\over (\bp+\bkappa_1)^2 }\Big\rangle 
+\Big\langle {1\over (\bp+\bkappa_2)^2 }\Big\rangle 
-{2\over \bp^2} \theta(\bp^2-\bkappa_1^2)\theta(\bp^2-\bkappa_2^2)\nonumber\\
&\Longrightarrow& {2 \over |\bp^2-\bkappa_1^2| +\delta^2} - 
{2\over \bp^2} \theta(\bp^2-\bkappa_1^2)\theta(\bp^2-\bkappa_2^2)\,,
\label{eq:7.2.7}
\eea
where we made a provision for the $\bkappa_1 \leftrightarrow \bkappa_2$ 
symmetry of the integrand.
At small $\bp$ the last term in (\ref{eq:7.2.7}) can safely be neglected
and the small-$\bp$ behaviour of $d\sigma^{(g2)}$ will be similar to
that of the spectrum of leading quarks (\ref{eq:6.3.9}) subject to
swapping the $g\to Q\bar{Q}$ splitting function for the $g\to gg$ one
as described in Appendix B. For the derivation of large-$\bp$ 
behaviour a more convenient representation is 
\bea
\Big\langle \Big\langle |\Psi(z,\bp+\bkappa_1)
-\Psi(z,\bp+\bkappa_2)|^2 \Big\rangle \Big\rangle &=&
{2 \bkappa_1^2 \over (\bp^2)^2}\theta(\bp^2-\bkappa_1^2) +
{2 \over \bkappa_1^2} \theta(\bkappa_1^2-\bp^2) \nonumber\\
&+&
{\tilde{K}}(\bp^2,\bkappa_1^2)+
{2\over \bp^2} \theta(\bp^2- \bkappa_1^2)\theta(\bkappa_2^2-\bp^2)\nonumber\\
\label{eq:7.2.8}
\eea
The logarithmic pole contribution is absorbed in 
${\tilde{K}}(\bp^2,\bkappa_1^2)$
which has the same properties as $K(\bp^2,\bkappa_1^2)$ in the expansion 
(\ref{eq:6.4.2.3}). Then the discussion of nuclear shadowing/antishadowing
properties of these three contributions will be entirely identical to 
that for the quark spectrum in section \ref{Sec:6.4}. Following the common
wisdom the logarithm $\log(\bp^2/\mu_G^2)$ in the resolved gluon
contribution, where $\mu_G$ is the appropriate
infrared cutoff for gluons, must be associated with the integrated 
density of gluons in the beam gluon, $G_B(z,\bp^2)$. The new item
in the expansion (\ref{eq:7.2.8}) is the last term; one can readily
verify that its contribution  is short of the $\log \bp^2$ and can be neglected 
compared to that from ${\tilde{K}}(\bp^2,\bkappa_1^2)$. Then, to the
leading $\log\bp^2$ approximation, 
\bea
{d\sigma_A^{(g2)} \over dz d^2\bp d\bb} &=& {d\sigma_N (g^*\to gg)
\over dz d^2\bp}\Biggr|_{res}
\cdot T(\bb)\cdot{2C_F \over C_A}\cdot{f_A^{(g)}(\bb,x_A,\bp^2) \over f(x_A,\bp^2)}
\nonumber\\
 &=& {d\sigma_N (g^*\to gg)
\over dz d^2\bp}\Biggr|_{res}\cdot 
T(\bb)\left[1+{C_A\over 2C_F}\Delta_{HT}(\bb,x_A,\bp)\right]
\label{eq:7.2.9}
\eea
The dominant contribution to (\ref{eq:7.2.6}) comes from the last,
nonshadowed, term in which the spectrum (\ref{eq:7.2.9}) is
convoluted with $\phi_g(\bb,x_A,\bkappa)$. The convolution effect
has been derived in section \ref{Sec:6.4}, see eq.~(\ref{eq:6.4.2.7}), the
generalization to our case gives the extra factor 
\beq
\rho_{A/N}^{(g)}(\bp) = 1+{C_A\over 2C_F}\Delta_{HT}(\bb,x_A,\bp)
\label{eq:7.2.10}
\eeq
and 
\bea
 &&{ d\Delta\sigma_A(g^* \to gg) \over dz d^2\bp d^2\bb} 
\Biggr|_{z\to 1} = {d\sigma_N (g^*\to gg)
\over dz d^2\bp}\Biggr|_{res}\cdot 
T(\bb)\cdot {C_A\over C_F}\Delta_{HT}(\bb,x_A,\bp)\, .
\label{eq:7.2.11}
\eea
A remarkable finding is that for this component of the gluon spectrum
the antishadowing Cronin
effect does not change from $z\ll 1$ to $z\to 1$. One can attribute 
that to the observation that although the nonlinear nuclear 
$k_{\perp}$-factorization in the two limits is of quite distinct 
form, the nonlinearity in the both cases is the same - a quadratic one.


\section{Gluon radiation off quarks and the spectrum of quarks from $q^* \to gq$}
\label{Sec:8}


\subsection{Linear $\bk_\perp$--factorization for gluon radiation off free
nucleons}
\label{Sec:8.1}

For completeness we show here the spectrum of gluons radiated from
a fast quark that propagated the nucleus, as well as the spectrum
of quarks after having shaken off their radiation cloud.

Here $a = q^*(\bb)$, $b =g(\bb_g)$, $c=q(\bb_q)$, where we also indicated 
the corresponding impact parameters. The lightcone momentum fraction
carried by the gluon is $z_g$. Because the transverse momentum 
distributions of produced gluons and scattered quarks are different,
for the sake of clarity we done them by $\bp_g$ and $\bp_q$, respectively. 

The relevant dipole distances are:
\beq \bb_g-\bb_q = \br \, , \, \bb'-\bb_q = z_g \br'\, ,\,
\bb_g-\bb_g' = \br-\br' \, ,\, \bb-\bb' = z_g(\br-\br') \, .
\label{eq:8.1.1}
\eeq

Again we start from the free nucleon case. The differential cross section is
written as
\bea {2 (2\pi)^2 d\sigma_N(q^* \to g(\bp_g) q) \over dz_g d^2\bp_g } &=&
\int d^2\br d^2\br' \exp[i\bp_g(\br-\br')] \Psi(z_g,\br) \Psi^*(z_g,\br')
\nonumber \\
& \times& \Big\{ \sigma_3(\br,z_g\br') + \sigma_3(\br',z_g\br) -
\sigma_{gg}(\br-\br') - \sigma_{q\bar{q}} (z_g(\br-\br')) \Big\}
\, ,
\nonumber\\
\label{eq:8.1.2}
\eea
and, again, now in momentum space:
\bea 
&&{2 (2\pi)^2 d\sigma_N(q^* \to g(\bp_g) q) \over dz_g d^2\bp_g }=\nonumber\\
&& \int d^2\bkappa f(x_A,\bkappa) \Big\{ {C_A \over 2 C_F}\Big[ |
\Psi(z_g,\bp_g) - \Psi(z_g,\bp_g+\bkappa)|^2 + |\Psi(z_g,\bp_g+\bkappa) -
\Psi(z_g,\bp_g+z_g \bkappa)|^2 \nonumber \\
&& - |\Psi(z_g,\bp_g) - \Psi(z_g,\bp_g + z_g \bkappa)|^2 \Big] + |\Psi(z_g,\bp_g) -
\Psi(z_g,\bp_g + z_g \bkappa)|^2 \Big\} \, .
\label{eq:8.1.3}
\eea
Which of course is nothing else but the
QCD--'Bethe-Heitler'--Bremsstrahlung--spectrum  \cite{Slava}
albeit in a not
so familiar representation in terms of lightcone wavefunctions.


\subsection{Nuclear case: nonlinear $\bk_\perp$ factorization for
gluon radiation}
\label{Sec:8.2}

We quote our result in the momentum space form:
\bea &&{ (2\pi)^2 d\sigma_A(q^* \to g(\bp_g) q) \over dz_g d^2\bp_g
d^2\bb} = \nonumber\\
&&S_{abs}(\bb) \int d^2\bkappa \phi(\bb,x_A,\bkappa)
\Big\{|\Psi(z_g,\bp_g) - \Psi(z_g,\bp_g+\bkappa) |^2
+ |\Psi(z_g,\bp_g+\bkappa) - \Psi(z_g,\bp_g+z_g \bkappa)|^2\Big\} \nonumber \\
&&+ \int d^2\bkappa_1 d^2\bkappa_2 \phi(\bb,x_A,\bkappa_1) \phi(\bb,x_A,\bkappa_2)
|\Psi(z_g,\bp_g+z_g \bkappa_1) - \Psi(z_g,\bp_g+\bkappa_1 + \bkappa_2)|^2 \, .
\label{eq:8.2.1}
\eea
Note the by now familiar decomposition into the absorbed 
linear and nonlinear - quadratic -terms.

Evidently, the spectrum of soft gluons does not depend on the
spin of the beam parton. Indeed, for soft gluons, $z_g \ll 1$,
the result (\ref{eq:8.2.1}) simplifies to
\bea
&&{ (2\pi)^2 d\sigma_A(q^* \to g(\bp_g) q) \over dz_g d^2\bp_g
d^2\bb}\Biggr|_{z_g \ll 1} = \nonumber\\
&&
\int d^2\bkappa  \Big\{ 2 S_{abs}(\bb)\phi(\bb,x_A,\bkappa) + \phi^{(2)}(\bb,x_A,\bkappa) \Big\}
|\Psi(z_g,\bp_g) - \Psi(z_g,\bp_g+\bkappa) |^2 \nonumber\\
&& =
\int d^2\bkappa \,
\phi_{gg}(\bb,x_A,\bkappa)
|\Psi(z_g,\bp_g) - \Psi(z_g,\bp_g+\bkappa) |^2\, ,
\label{eq:8.2.2}
\eea
which is a precise counterpart of eq.~(\ref{eq:7.1.9}) for the $g^*\to gg$
breakup. In the large-$N_c$ approximation $\phi_g(\bb,x_A,\bkappa)=
\phi(\bb,x_A,\bkappa )$ and $\Big(\phi \otimes \phi\Big)(\bb,x_A,\bkappa)=
\phi_{gg}(\bb,x_A,\bkappa)$.

Consequently, the antishadowing Cronin effect for slow gluons from $q^*\to qg$
will be identical to that from $g^*\to gg$,
\bea
\Delta_{gq}(z_g\ll 1; x_A,\bp_g)=\Delta_{gg}(z_g\ll 1;x_A,\bp_g)=
2\Delta_{Qg}(z_Q\ll 1;x_A,\bp_g)\, .
\label{eq:8.2.3}
\eea
For the numerical estimates for $pAu$ collisions 
one must take $\Delta_{Cronin}(x_A,\bp_g)$ for $j=8$ in Figs.
\ref{fig:DeltaCronin1}, \ref{fig:DeltaCronin2}.
On the other hand a comparison of  eqs. (\ref{eq:8.2.1}) and (\ref{eq:5.5})
shows that the spectrum of leading gluons with $z_g\to 1$
from $q^*\to qg$ will have precisely the same convolution
form (\ref{eq:6.1.4}) as the spectrum of leading quarks from $g^* \to Q\bar{Q}$
which in view of the findings of sections \ref{Sec:6.4} and \ref{Sec:7.2} entails 
\bea
\Delta_{gq}(z_g\to 1; x_A,\bp_g)= 
\Delta_{Qg}(z_Q\to 1;x_A,\bp_g)=\Delta_{gg}(z_g\to 1; x_A,\bp_g).
\label{eq:8.2.4}
\eea
This finding suggests that the equality of antishadowing Cronin effect for leading and
soft gluons does not depend on the beam parton be it a quark or a gluon.


\subsection{$\bp_q$--spectrum of quarks from the breakup
$q^* \to gq$ on a free nucleon}
\label{Sec:8.3}

We now turn to our last application, the spectrum of quarks from
the excitation of the $qg$--Fock state of a fast quark. In the
practically interesting case of forward quark jets in $pA$ collisions
at RHIC the beam quarks $q^*$ have small transverse momentum and
the production of high-$\bp$ quarks is not affected 
by virtual radiative corrections,
see the discussion in Sect. \ref{Sec:7.2}.
Without further details we quote again first the dipole
representation:
\bea
{2 (2\pi)^2 d\sigma_N(q^* \to q(\bp_q) g) \over dz_q d^2\bp_q} =
\int d^2\br d^2\br' \Psi(z_q,\br) \Psi^*(z_q,\br') \exp[i \bp_q(\br-\br')]
\nonumber \\
\{ \sigma_3(\br , \br- z_q \br') + \sigma_3(\br',\br'-z_q\br) -
\sigma_{q\bar{q}}(\br-\br') - \sigma_{q\bar{q}}(z_q(\br-\br'))\}
\, \, , 
\label{eq:8.3.1}
\eea
while the linear $k_\perp$--factorization result reads
\bea
&&{2 (2\pi)^2 d\sigma_N(q^* \to q(\bp_q) g) \over dz_q d^2\bp_q} =
\int d^2\bkappa f(x_A,\bkappa)\nonumber\\
&& \Big\{ {C_A \over 2 C_F} \Big[
|\Psi(z_q,\bp_q) -\Psi(z_q,\bp_q+\bkappa)|^2 \nonumber \\
&&+ |\Psi(z_q,\bp_q)
-\Psi(z_q,\bp_q + z_q \bkappa) |^2 - | \Psi(z_q,\bp_q+\bkappa) - \Psi(z_q,\bp_q+z_q
\bkappa)|^2 \Big ]\nonumber \\
&&+ |\Psi(z_q,\bp_q+\bkappa) - \Psi(z_q,\bp_q + z_q \bkappa)|^2 \Big\} \, .
\label{eq:8.3.2}
\eea
A closer comparison of the $\bp$ spectra of gluons and quarks
from $g^* \to gq$ reveals interesting differences which
hold in both the nonabelian and abelian, $C_A \to 0$, cases.
Evidently, in the $\bp$--integrated 
spectra we would obtain the longitudinal spectrum of quarks simply 
by swapping $z_g \to 1-z_g = z_q$ in the formula for the gluon 
spectrum and vice versa. The similar procedure for the $\bp$-spectra,
as advocated in \cite{KopSchafTar} would be entirely wrong.


\subsection{Quark spectrum from the breakup of the $qg$--Fock state
of a fast quark on a nuclear target}
\label{Sec:8.4}

Routine application of the above described technique yields
\bea 
&&{ (2\pi)^2 d\sigma_A(q^* \to q(\bp_q) g) \over dz_q d^2\bp_q
d^2\bb} = \nonumber\\
&&S_{abs}(\bb) \int d^2\bkappa \phi(\bb,x_A,\bkappa)
\Big\{|\Psi(z_q,\bp_q) - \Psi(z_q,\bp_q+\bkappa) |^2
+ |\Psi(z_q,\bp_q ) - \Psi(z_q,\bp_q+z_q \bkappa)|^2\Big\} \nonumber \\
&&+ \int d^2\bkappa_1 d^2\bkappa_2 \phi(\bb,x_A,\bkappa_1) \phi(\bb,x_A,\bkappa_2)
|\Psi(z_q,\bp_q+ \bkappa_1) - \Psi(z_q,\bp_q + z_q \bkappa_2)|^2 \, . 
\label{eq:8.4.1}
\eea
As usual, the absorbed linear $k_{\perp}$-factorization term
is the counterpart of the large-$N_c$ free-nucleon spectrum
(\ref{eq:8.3.2}).
Note also that the $\bp_q$-spectrum of slow quark jets from $q^*\to qg$
coincides with the spectrum (\ref{eq:6.1.3}) form $g^* \to Q\bar{Q}$
and the Cronin effect for the two cases will be identical:
\beq
\Delta_{qq}(z_q\ll 1; x_A,\bp_q)=
\Delta_{Qg}(z_Q\ll 1;x_A,\bp_q)\, .
\label{eq:8.4.2}
\eeq
For the numerical estimates for $pAu$ collisions 
one must take $\Delta_{Cronin}(x_A,\bp_q)$ for $j=4$ in Figs.
\ref{fig:DeltaCronin1}, \ref{fig:DeltaCronin2}.

The spectrum  of leading-quark jets, $z_q \to 1$,
equals 
\bea { (2\pi)^2 d\sigma_A(q^* \to q(\bp_q) g) \over dz_q d^2\bp_q
d^2\bb}\Big|_{z_q \to 1} = 2 \, S_{abs}(\bb) \int d^2\bkappa
\phi(\bb,x_A,\bkappa) |\Psi(z_q,\bp_q) - \Psi(z_q,\bp_q+\bkappa) |^2
\nonumber \\
+ \int d^2\bkappa_1 d^2\bkappa_2 \phi(\bb,x_A,\bkappa_1) \phi(\bb,x_A,\bkappa_2)
|\Psi(z_q,\bp_q+ \bkappa_1) - \Psi(z_q,\bp_q + \bkappa_2)|^2 \, , \nonumber\\
\label{eq:8.4.3}
\eea
where the second term is similar to $d\sigma^{(2g)}$ of eq. (\ref{eq:7.13})
which emerged in the discussion of the spectrum of leading gluons 
from $g^* \to gg$. Its antishadowing properties have been studied 
in section \ref{Sec:7.2} and entail
\bea
\Delta_{qq}(z_q\to 1; x_A,\bp_q)=\Delta_{qq}(z_q\ll 1; x_A,\bp_q)
\label{eq:8.4.4}
\eea
The antishadowing effect for leading quarks from $q\to qg$ is one half of that for 
leading gluons from $g\to gg$.

The above derived $z_q$-distribution quantifies the energy loss for 
the radiation of gluons, i.e., gives a solution of the LPM problem
for finite transverse momentum $\bp$ of the observed quark jet. The case of
quark jets is important for its relevance to the large
pseudorapidity, high-$\bp$
hadron production in deuteron-gold collisions at RHIC. The
numerical studies of the LPM effect contribution to nuclear quenching 
of forward high-$\bp$
hadrons observed experimentally by the BRAHMS collaboration \cite{BRAHMS}
will be reported elsewhere.

\section{Discussion of the results}
\label{Discussion}

Extending the technique developed earlier
\cite{Nonlinear}
we elucidated the pattern of $k_{\perp}$--factorization breaking
for the single gluon and quark jet production from $g^*\to Q\bar{Q},
\,g^*\to gg,\, q^*\to qg$ breakup on nuclear targets. In all
the cases we derived an explicit nonlinear, from quadratic to
cubic, $k_{\perp}$-factorization
representations for the inclusive  single-jet transverse 
and longitudinal momentum
spectra in terms of the collective nuclear unintegrated gluon
density. The results are simple and elegant and have not been
presented in this form in the existing literature on
the subject. They also give a solution of the LPM problem
for finite transverse momentum $\bp$ of the observed secondary quark 
and gluon jet. One of the evident future applications will be
to the nuclear quenching of forward high-$\bp$
hadrons observed experimentally by the BRAHMS collaboration \cite{BRAHMS}.

The  unintegrated collective nuclear gluon
density which defines the nuclear observables is shown to
change from the quark to gluon observables even within the
same process what exemplifies the point that an ultrarelativistic
nucleus can not be described by a single gluon density, the
collective nuclear density must rather be described by
a density matrix in color space \cite{Nonlinear}.
The often used linear $k_{\perp}$-factorization
description of particle production off nuclei 
(e.g. \cite{KharLevMcL}, for the recent applications and more
references see also
\cite{BaierKovWied,JalilianRaju,GelisRaju,Kharzeev:2003sk}) 
is definitely not borne out by the multiple scattering theory.
There is a striking
contrast to the single-quark jet transverse momentum spectrum
from $\gamma^*\to q\bar{q}$ breakup on a nucleus for which linear
$k_{\perp}$-factorization is fulfilled, and our analysis shows
that that the linear $k_{\perp}$-factorization in $\gamma^*\to q\bar{q}$
is rather an exception due to the color-singlet nature of the
photon by which the single-jet spectrum accidentally
becomes an abelian problem \cite{Nonlinear}.

Although the linear $k_{\perp}$-factorization for
hard processes in nuclei does not exist,
for all breakup processes we identified the `abelian' regime in
which the transverse momentum spectra of soft quark and gluons jets
with $z_{q,g}\ll 1$ take the process-dependent
linear $k_{\perp}$-factorization
form. Specifically, the production of quark jets is described
by the collective
gluon density $\phi(\bb,x_A,\bkappa)$
defined for the triplet-antitriplet color dipole.
The production of gluon jets is described in terms
of the collective gluon density $\phi_{gg}(\bb,x_A,\bkappa)$ defined for
the octet-octet color dipole. 
The color-dipole representation
for such a slow gluon limit coincides with that derived before  
for the slow gluon production by Kovchegov and Mueller 
\cite{KovchegovMueller}. The 
reason behind the recovery of the Kovchegov-Mueller form for 
slow jets is that in this limit the single-parton spectrum
becomes an abelian problem \cite{Nonlinear}. Such a form 
is of limited practical significance,
though. Evidently, the limit
of $z_q \ll 1$ is irrelevant to the open charm
production, in the gluon jet production too
the dominant contribution to the 
observed jet cross
section will come from finite $z$ to be described 
by our nonlinear $k_{\perp}$-factorization. The full
$z$-dependence is also mandatory for the evaluation
of the LPM quenching of forward high-$\bp$ jets.
The production of gluons
from $g\to g_1g_2$ has been discussed also in 
\cite{KopNemSchafTar}, their equation for the nuclear spectrum
contradicts our general result (\ref{eq:3.3}). The earlier
work on $q\to qg$ by  \cite{KopSchafTar} starts
with the two-parton inclusive spectrum. When compared to our
general equation (\ref{eq:3.3}), in ref.
\cite{KopSchafTar} the coupled-channel
S-matrix $S_{q\bar{q}gg}^{(4)}$ has illegitimately been 
substituted by a single-channel Glauber exponential. Despite 
this inconsistency, the final  color dipole representation 
of Ref. \cite{KopSchafTar} for the 
single-parton spectrum agrees with ours.

Recently, the 
production of heavy quarks from the excitation $q \to q'g $,
$g \to Q\bar{Q}$ has been discussed in the color dipole framework 
in ref. \cite{Tuchin:2004rb}. 
We have however not been able 
to reconcile the results given there 
with our findings on the cancellation 
of spectator interactions. 
Very recent works  \cite{GelisRaju,Blaizot:2004wu} start with the
reference to the color glass condensate model, but in actual
applications the Glauber-Gribov multiple
scattering theory is used and a quantity similar to 
the collective unintegrated glue of \cite{Nonlinear,NSSdijet} is introduced.
Although no explicit formulas corresponding to
eq.(\ref{eq:5.5}) for the heavy quark cross section 
are given, the authors are aware in part of our criticism \cite{Nonlinear}
of the applicability of linear $k_\perp$ factorization in
the saturation regime.

Regarding the applicability domain of our formalism
the considered processes correspond to production of 
parton pairs with $x_{b,c} \lsim x_A \sim 10^{-2}$, where 
$x_{b,c}$ are defined with respect to the nucleus. In the
$pA$ collisions at RHIC this requires that both partons 
were produced in the proton hemisphere.
The
consistent discussion of higher order processes
and of the emerging BFKL evolution of nuclear 
spectra remains an open problem. 
Generally speaking one will have to address 
a situation in which the coherence of more than two partons 
over the nucleus is important, and the question is 
in how far the rescattering 
of partons faster/slower than the measured jet is reabsorbed
into the (non-)linear evolution of the beam/target 
effective gluon densities. Such evolution effects 
would perhaps not feature too prominently at RHIC energies, 
where the rapidity span overlaps well with the applicability
region of our approach. 
Still such studies will be important to establish the viability
of our approach to higher energies such as LHC. 
They would also sharpen our understanding of the similarities
and differences of our color dipole formalism and
the Color Glass Condensate approach (for recent explorations of the 
energy dependence of the Cronin effect for slow gluon production
within those models, see e.g. 
\cite{ BaierKovWied,JalilianRaju,Kharzeev:2003wz,
Albacete,Blaizot:2004wu,Iancu:2004bx,Kharzeev:2004yx}).

The most obvious application of the derived nonlinear 
nuclear $k_{\perp}$-factorization is to nuclear effects 
in the single-jet spectrum which are usually referred to
as the Cronin effect. Our derivation 
unambiguously relates it to the antishadowing property
of the collective nuclear gluon density \cite{Nonlinear,NSSdijet}. 
The Cronin effect was shown to exhibit interesting variations across the 
phase space. 
Based on the numerical studies
of the antishadowing effect in the collective nuclear gluon
density \cite{NSSdijet} we presented numerical estimates for
the Cronin effect.  One feature of these estimates is noteworthy:
The realistic models for the unintegrated gluon
density of the proton \cite{INDiffGlue} suggest a somewhat 
small value of the saturation scale, $Q_A \approx 1$ GeV, even
for the heaviest nuclei. Still, the Cronin effect is found to reach
its peak value at $|\bp| \approx (2\div2.5)$ GeV, which is well within
the pQCD domain.

To summarize, our analysis establishes nonlinear $k_{\perp}$-factorization
as a universal feature of hard production off nuclear targets.
The presented formalism can readily be extended to the nuclear
dependence of jet-jet correlations and nuclear quenching
of forward quark jets, the corresponding work is in progress.
  
\noindent
{\bf Acknowledgements:} We are grateful to B.G. Zakharov for illuminating
discussions and suggestions.

\section*{Appendix A: The derivation of $S^{(n)}$}
\setcounter{equation}{0}
\renewcommand{\theequation}{A.\arabic{equation}}

Here we expose some of the technicalities behind the  
derivation of the master equation (\ref{eq:2.5}) and
the reduction of the four-parton $S$-matrix to the two-body
$S$-matrices in equation (\ref{eq:2.7}).
The color-dipole properties of excitation amplitudes are 
elucidated and the calculation of the $S$-matrices $S^{(n)}$ is greatly 
simplified in the $q\bar{q}$ representation of the gluon interactions.
We demonstrate the principal virtues of this technique on an
example of $g^*\to Q\bar{Q}$ on a free-nucleon target, for the 
extension to a nuclear target see \cite{Nonlinear}.
In the color space the $g_a Q\bar{Q}$
vertex contains the $SU(N_c)$ generator $t^a$, so that in the quark color
space the contribution of the diagram of fig. \ref{fig:single-jet_a_to_bc}b is proportional to
$ S(\bb_b)t^a S^{\dagger}(\bb_c)$, whereas the contribution of the 
diagram of fig. \ref{fig:single-jet_a_to_bc}c is proportional to $-t^d S_{da}(\bb)$. Here
$S^{\dagger}(\bb_c)$ describes the interaction of the antiquark, and 
$S_{da}(\bb) =\langle g_d|S_g(\bb)|g_a\rangle$ describes the
transition between the two gluons in color states $a$ and $d$.
Consequently, the properly normalized excitation operator is
\beq
{\cal M}_a(\bb_b,\bb_c,\bb) = {1\over \sqrt{C_F N_c}}
 \Big( S(\bb_b)t^a S^{\dagger}(\bb_c)- t^d S_{da}(\bb)\Big)
\label{eq:A.1}
\eeq
where $C_F N_c = {\rm Tr \,} t^a t^a$. 
The above vertex operators $t^a$ have been suppressed in the expansion
(\ref{eq:2.3}). 

Now we note that 
\beq
S_{da}(\bb) =
{1\over T_F} {\rm Tr \,}\Big( t^{d}S(\bb)t^{a}S^{\dagger}(\bb)\Big)
\label{eq:A.2}
\eeq
and, upon the Fierz transformation,
\beq
t^d S_{da}(\bb) = S(\bb)t^a S^{\dagger}(\bb)\,,
\label{eq:A.3}
\eeq
which leads to a particularly simple form of the excitation
operator:
\beq
{\cal M}_a(\bb_b,\bb_c,\bb) = {1\over \sqrt{C_F N_c}}
 \Big( S(\bb_b)t^a S^{\dagger}(\bb_c)- S(\bb)t^a S^{\dagger}(\bb)\Big)
\label{eq:A.4}
\eeq
Note how ${\cal M}_a(\bb_b,\bb_c,\bb)$ vanishes if $\bb_b=\bb_c =\bb$:
in this limit the $Q\bar{Q}$ behaves as a pointlike color-octet state
indistinguishable from the gluon and its $Q\bar{Q}$ structure can not
be resolved. What enters the integrand of (\ref{eq:2.5}) is
${\rm Tr \,}\Big( {\cal M}_a^{\dagger}(\bb_b',\bb_c',\bb'){\cal M}_a(\bb_b,\bb_c,\bb)\Big)$.

Before proceeding further we notice that the $S$-matrix for the 
color singlet $q\bar{q}$ dipole equals
\beq
S_2(\bb_b - \bb_c) = {1 \over N_c} {\rm Tr \,} \Big(S(\bb_b)S^{\dagger}(\bb_c)\Big)
=1-\Gamma_2(\bB,\bb_b - \bb_c)\,,
\label{eq:A.5}
\eeq
where $\br = \bb_b - \bb_c$ is the relevant dipole size and 
$\Gamma_2(\bB,\bb_b - \bb_c)$ is the corresponding profile function.
Upon the integration over the dipole impact parameter, $\bB$, one finds
the color-dipole cross section
\beq
\sigma_{q\bar{q}}(x,\br) = 2\int d^2\bB \Gamma_2(\bB,\br)\, .
 \label{eq:A.6}
\eeq 
The excitation on a free nucleon target is described to the single-gluon 
exchange approximation. In this approximation
\beq
S_2^2(\bb_b - \bb_c) =1-2\Gamma_2(\bB,\bb_b - \bb_c)\,,
\label{eq:A.7}
\eeq

Now we can identify the $S^{(n)}$ in the integrand of (\ref{eq:2.5}) and
give explicit expressions in terms of $\Gamma_2(\bB,\br)$. Making use 
of 
\beq
{\rm Tr \,}\Big( t^a A t^a B\Big) = {1\over 2} \Big( {\rm Tr \,} \Big(A\Big) \cdot  {\rm Tr \,} \Big(B \Big)
- {1\over N_c} {\rm Tr \,} \Big(AB\Big) \Big)
\label{eq:A.8}
\eeq
and applying the unitarity condition (\ref{eq:2.6**}) wherever appropriate
one readily finds
\bea
S^{(2)}(\bb,\bb') &=& {1\over C_F N_c} {\rm Tr \,}\Big(
 S(\bb')t^a S^{\dagger}(\bb')S(\bb)t^a S^{\dagger}(\bb)\Big) 
= 1-{C_A \over C_F} \Gamma_2(\bB,\bb'-\bb)\\
\label{eq:A.8.1}
S^{(3)}(\bb_b,\bb_c,\bb') &=& {1\over C_F N_c} {\rm Tr \,}\Big(
 S(\bb_b)t^a S^{\dagger}(\bb_c)S(\bb')t^a S^{\dagger}(\bb')\Big) \\  
\label{eq:A.8.2}
= 1 &-&\left\{ {C_A \over 2C_F} \Big(\Gamma_2(\bB,\bb'-\bb_b) +
\Gamma_2(\bB,\bb'-\bb_c)\Big) - {1\over N_c^2-1} \Gamma_2(\bB,\bb_b - \bb_c) \right\}\nonumber\\
\label{eq:A.8.3}
S^{(3)}(\bb_b',\bb_c',\bb) &=& {1\over C_F N_c} {\rm Tr \,}\Big(
 S(\bb_b')t^a S^{\dagger}(\bb_c')S(\bb)t^a S^{\dagger}(\bb)\Big)\nonumber\\
=  1 &-&\left\{ {C_A \over 2C_F} \Big(\Gamma_2(\bB,\bb-\bb_b') +
\Gamma_2(\bB,\bb -\bb_c')\Big) - {1\over N_c^2-1} \Gamma_2(\bB,\bb_b' - \bb_c') \right\}\,.\nonumber\\
\label{eq:A.8.4}
\eea
In the $S^{(3)}$ one readily identifies the profile functions of the $q\bar{q}g$ cross
section as introduced in \cite{NZ94}. The case of $S^{(4)}$ is more tricky:
\bea
S^{(4)}(\bb_b,\bb_c,\bb_b',\bb_c') &=& {1\over C_F N_c} {\rm Tr \,}
\Big( S(\bb_b)t^a S^{\dagger}(\bb_c)S(\bb_c')t^a S^{\dagger}(\bb_b')\Big) \nonumber\\
&=& {1\over N_c^2 -1} \Big\{N_c^2\Big(1- \Gamma_2(\bB,\bb_c -\bb_c')-
\Gamma_2(\bB,\bb_b -\bb_b')\Big) \nonumber\\
&-& {1\over  N_c} {\rm Tr \,}\Big[
 S(\bb_b)S^{\dagger}(\bb_c)S(\bb_c') S^{\dagger}(\bb_b')\Big] \Big\} \, .
\label{eq:A.10}
\eea
The last term in (\ref{eq:A.10}) is an operator in the $q\bar{q}$ color
space and in general $S^{(4)}$ is the coupled-channel operator
derived in \cite{Nonlinear}. In the case of the single-particle particle
spectrum
$\bb_c'=\bb_c$ and the straightforward application of the unitarity
condition (\ref{eq:2.6**}) yields
\beq
S^{(4)}(\bb_b,\bb_c,\bb_b',\bb_c) = S_2(\bb_b -\bb_b')= 1- \Gamma_2(\bB,\bb_b -\bb_b')\,.
\label{eq:A.11}
\eeq
As quoted above the results are for the free nucleon target, for the nuclear
target one must apply the Glauber--Gribov exponentiation to the
relevant $S^{(n)}$.

\section*{Appendix B:}
\setcounter{equation}{0}
\renewcommand{\theequation}{B.\arabic{equation}}

In all the transverse momentum spectra the recurrent quantity 
is $|\Psi(z,\bp)-\Psi(z,\bp-\bkappa)|^2$. This quantity for the incident photon  
is found in \cite{NZ91,Nonlinear}. For transverse photons
and flavor $f$, i.e., for the $\gamma^* \to Q\bar{Q}$ excitation, 
\bea
|\Psi(z,\bp)-\Psi(z,\bp-\bkappa)|^2=
2N_c e_Q^2\alpha_{em}\nonumber\\
\times \left\{[z^{2}+(1-z)^{2}]
\left({\bp \over  \bp^{2}+\varepsilon^{2}} -
{\bp-\bkappa  \over  (\bp-\bkappa )^{2}+\varepsilon^{2}}\right)^2
\right.\nonumber\\
\left. +m_{Q}^{2}
\left({1   \over  \bp^{2}+\varepsilon^{2}}-
{1 \over  (\bp-\bkappa )^{2}+\varepsilon^{2}}\right)^2
\right\}\,, 
\label{eq:B.1}
\eea
The term $ \propto m_Q^2$ must only be kept for heavy quarks.
The same quantity for the $g^* \to Q\bar{Q}$ is obtained from
(\ref{eq:B.1}) by the substitution $N_c e_Q^2\alpha_{em} \to T_F\alpha_S(Q_a^2)$
and the substitution in (\ref{eq:B.2})of the virtuality of the photon $Q^2$ by the
virtuality of the beam gluon $Q_{g^*}^2 $.
For the general case $a \to bc$
\beq
\varepsilon^2 = z_b z_c Q_a^2 + z_b m_c^2 + z_c m_b^2 \,,
\label{eq:B.2}
\eeq
where $Q_a^2$ is the virtuality of parton $a$ 
and the mass dependent term is important for heavy flavours.
Now note that the factor $T_F[z^{2}+(1-z)^{2}]$ which emerges in the
first term in the rhs of (\ref{eq:B.1}) is precisely the familiar splitting
function $P_{Qg}(z)$. For all other cases  $|\Psi(z,\bp)-\Psi(z,\bp-\bkappa)|^2$
is obtained from that for $g^* \to Q\bar{Q}$ by the substitution of 
$P_{Qg}(z)$ by the real-emission part of the 
relevant splitting function for $z<1$ found in all textbooks 
\cite{Textbook}. Take for instance the fragmentation of light quarks 
$q\to qg$. The 
$|\Psi(z_q,\bp)-\Psi(z_q,\bp-\bkappa)|^2$ which enters the spectrum
of quarks must be evaluated with the splitting function
\beq
P_{qq}(z_q)=C_F {1+z_q^2 \over (1-z_q)}\,.
\label{eq:B.2.1}
\eeq
The  $|\Psi(z_g,\bp)-\Psi(z_g,\bp-\bkappa)|^2$ which enters the spectrum
of gluons in the same process must be evaluated with
\beq
P_{gq}(z_g)=C_F {1+(1-z_g)^2 \over z_g} \, .
\label{eq:B.2.2}
\eeq
In the fragmentation of gluons $g \to gg$ one must 
take $|\Psi(z,\bp)-\Psi(z,\bp-\bkappa)|^2$
with the splitting function
\beq
P_{gg}(z)=2C_A \left[{1-z \over z} + {z \over 1-z } + z(1-z) \right]\,.
\label{eq:B.2.3}
\eeq

If $\varepsilon^2$ is negligible small compared to $\bp^2$, then one can
use the large-$\bp$ approximation:
\beq
\left({\bp \over  \bp^{2}} -
{\bp-\bkappa  \over  (\bp-\bkappa )^{2}}\right)^2 = {\bkappa^2
\over \bp^{2}(\bp-\bkappa )^{2}} \, ,
\label{eq:B.3}
\eeq
{and it is worth to recall the emerging exact factorization 
of longitudinal and transverse momentum 
dependencies which is a well known feature of the high energy limit.}

\end{document}